\documentclass[pre,floatfix,a4paper,nofootinbib,amssymb,amsmath,showpacs,superscriptaddress]{revtex4}
\usepackage{graphicx}
\usepackage{bm}
\usepackage{eucal}
\usepackage{mathrsfs}

\newcommand {\be} {\begin{equation}}
\newcommand {\ee} {\end{equation}}
\newcommand {\Be}{\begin{eqnarray*}}
\newcommand {\Ee} {\end{eqnarray*}}
\newcommand {\bey} {\begin{eqnarray}}
\newcommand {\eey} {\end{eqnarray}}
\newcommand{\bit}{\begin{itemize}}      
\newcommand{\eit}{\end{itemize}}
\newcommand{\bfl}{\begin{flusleft}}
\newcommand{\efl}{\end{flusleft}}
\newcommand{\bfr}{\begin{flushright}}
\newcommand{\bc}{\begin{center}}
\newcommand{\ec}{\end{center}}
\newcommand{\ben}{\begin{enumerate}}    
\newcommand{\een}{\end{enumerate}}

\newcommand{\comment}[1]{}

\newcommand{\E}{\mathrm{e}}

\newcommand{\p}[1]{\left({#1}\right)}
\newcommand{\pq}[1]{\left[{#1}\right]}

\newcommand{\D}{\mathrm{d}}

\newcommand{\average}[1]{\left\langle{#1}\right\rangle}

\begin{document} 

\title{Free energy landscape of mechanically unfolded model proteins: \\
extended Jarzinsky versus inherent structure reconstruction}

\author{Stefano Luccioli}
\email{stefano.luccioli@fi.isc.cnr.it}
\affiliation{Istituto dei Sistemi Complessi, CNR, via Madonna del Piano 10, I-50019 Sesto Fiorentino, Italy}
\affiliation{INFN, Sez. Firenze, and CSDC, via Sansone, 1 - I-50019 Sesto Fiorentino, Italy}
\author{Alberto Imparato}
\email{alberto.imparato@polito.it}
\affiliation{ISI Foundation, Viale Settimio Severo 65,
  Villa Gualino, I-10133 Torino, Italy}
\author{Alessandro Torcini}
\email{alessandro.torcini@cnr.it}
\affiliation{Istituto dei Sistemi Complessi, CNR, via Madonna del Piano 10, I-50019 Sesto Fiorentino, Italy}
\affiliation{INFN, Sez. Firenze, and CSDC, via Sansone, 1 - I-50019 Sesto Fiorentino, Italy}
\affiliation{Centre de Physique Th{\'e}orique, Campus de Luminy, 13288 Marseille, France}

\begin{abstract}
The equilibrium free energy landscape of off-lattice model heteropolymers as a
function of an internal coordinate, namely the end-to-end distance, is 
reconstructed from out-of-equilibrium steered molecular dynamics data.
This task is accomplished via two independent methods: by employing an extended 
version of the Jarzynski equality (EJE) and the inherent structure (IS) formalism.
A comparison of the free energies estimated with these two schemes with equilibrium results
obtained via the umbrella sampling technique reveals a good quantitative agreement 
among all the approaches in a range of temperatures around the ``folding transition''
for the two examined sequences. In particular, for the sequence with good 
foldability properties, the mechanically induced structural 
transitions can be related to thermodynamical aspects of folding. Moreover,
for the same sequence the knowledge of the landscape profile allows for a good 
estimation of the life times of the native configuration for
temperatures ranging from the folding to the collapse temperature.
For the random sequence, mechanical and thermal unfolding
appear to follow different paths along the landscape.
\end{abstract}


\maketitle

\section{Introduction}\label{one}

Several states of matter are characterized by a non trivial free energy 
landscape (FEL), which can be at the origin of peculiar structural 
and  dynamical features.
Supercooled liquids, glasses, atomic clusters and biomolecules \cite{wales} are
typical examples of systems whose thermodynamical behavior can be
traced back to the intricate topological properties of the underlying FEL. 
The pioneering work by Stillinger and Weber on inherent structures (ISs) of liquids \cite{still2}
revealed the importance of investigating the stationary points of the potential energy
surface (PES) for characterizing their dynamical and thermodynamical properties. Similar
approaches have been proposed and successfully applied, in glasses~\cite{sastry} and supercooled
liquids~\cite{angelani}, to the identification of the structural--arrest temperature.
This temperature marks a topological transition from a dynamics evolving in a landscape 
dominated by minima to one where unstable saddles play a major role~\cite{angelani, grigera}.

More recently, this kind of analysis has been applied to the study of protein
models \cite{thiru_1,wales,wales_arti,krivov,evans,baum,nakagawa,kim-keyes}. 
In particular, several studies have been devoted to the reconstruction 
of the PES and of the FEL topology in terms of graphs (at various levels of coarse
graining) connecting the folded states to the unfolded structures~\cite{wales_arti,krivov,evans}.
The knowledge of the graph structure connecting the various metastable states and
of the probability transitions among them allows for a reconstruction of the folding
dynamics in terms of a master equation~\cite{wales2004}. Moreover, detailed analysis of the thermodynamical 
and dynamical features, characteristics of proteins, have been quite recently carried out
in terms of ISs \cite{baum,nakagawa,kim-keyes}.  These analysis
suggest that the folding process of a protein towards its native configuration 
depends crucially on the structure and topological properties of its (free) energy 
landscape. Confirming somehow the conjecture that the FEL of a protein
has a funnel--like shape: the native configuration being located
inside the so--called native valley at the bottom of the funnel itself \cite{funnel}.

On the other hand, mechanical unfolding of single biomolecules represents a powerful technique
to extract information on their internal structure as well as on
their unfolding and refolding pathways \cite{exp_pap,ubiquitin, brockwell,ubi_sim,imp04}.
However, mechanical unfolding of biomolecules is an out-of-equilibrium
process: unfolding events occur on time scales much shorter than the
typical relaxation time of the molecule towards equilibrium.
Nonetheless, by using the equality introduced by Jarzynski \cite{jarzynski}, the
free energy of mechanically manipulated biomolecules can be recovered as a function
of an externally controlled parameter \cite{ritort,west_paci}. 
Moreover, an extended version of the Jarzynski equality (EJE)
has been proposed in order to estimate the equilibrium free energy 
landscape in absence of applied forces 
as a function of an internal coordinate of the system 
(usually, the end-to-end distance $\zeta$)~\cite{HumSza,Seif,noi2,noi3}.
Quite recently this approach has been successfully applied to data
obtained from nanomanipulation of titin I27 domain with 
Atomic Force Microscopy (AFM)~\cite{new,ISV} 
and from steered molecular dynamics simulations of a mesoscopic off-lattice protein 
model \cite{nostro}. The analysis reported in \cite{nostro} was devoted to
a single sequence previously identified as a reasonably fast folder
\cite{guo_thir,guo} and it was essentially performed at an unique temperature.

As an extension of the analysis performed in Ref. \cite{nostro},
in the present paper we reconstruct, for two different sequences with bad
and good folding properties, the equilibrium FEL as a function of the 
end-to-end distance $\zeta$ in two distinct ways: namely, by employing 
the EJE approach and the IS distributions. We will show that 
specific features of the landscapes characteristic of a protein, 
i.e. a good folder, can be singled out from a comparison of the two 
approaches. Furthermore, the different unfolding structural transitions
can be associated to the detachment of specific strands of the examined heteropolymers.
In particular, the investigation of the IS distributions allows us to give an estimate
of the energetic and entropic barriers separating the native state from the completely stretched configuration. 
Moreover, for the good folder the temperature dependence of the free energy barrier heights
and the unfolding times can be related.

An important aspect to clarify is the relationship between the 
thermal and mechanical unfolding pathways of proteins: 
experimental \cite{exp_term_pul} as well as numerical works \cite{paci_karplus,west_paci_2} 
seem to suggest that these paths are indeed different. However, there are
indications that the thermal paths can be recovered also via the manipulation procedure
in the limit of very low pulling velocities~\cite{west_paci_2}.
This seems to be in agreement with our findings for the good folder, 
which indicate that the observed structural transitions, 
induced by mechanical unfolding, can be put in direct relationship with the thermal 
transitions usually identified for the folding/unfolding process.

The paper is organized as follows, Sect.~\ref{two} is devoted
to the introduction of the employed model and sequences, as well 
as of the simulation protocols. In Sect.~\ref{three_I} it is
explained how to combine the umbrella sampling technique \cite{umbrella} with 
the weighted histogram analysis method \cite{wham} in order to recover
the equilibrium free energy profile as a function of 
an internal coordinate of the system. The inherent structure formalism and
the extended Jarzynski equality are briefly illustrated in Sect.~\ref{three} 
and Sect.~\ref{four}, respectively. The thermodynamical properties of the
studied sequences are reported in Sect.~\ref{five}. While Sect.~\ref{six}
(resp. Sect. \ref{seven}) is devoted to the free energy landscape reconstruction 
in terms of the extended Jarzinsky equality (resp. inherent structure approach).
In Sect.~\ref{seven} the two methods are also compared and discussed.
Finally, the results are summarized in Sect.~\ref{eight}.

\section{Model and simulation protocol}\label{two}

\subsection{The model}
\label{model}

The model studied in this paper is a modified version of the 3d off-lattice
model introduced by Honeycutt-Thirumalai \cite{honey} and successively 
generalized by Berry {\it et al.} to include a harmonic interaction between
next-neighbouring beads instead of rigid bonds \cite{berry}.
This model has been widely studied in the context of thermally driven
folding and unfolding \cite{honey,guo_thir,guo,berry,veit,evans,kim,kim-keyes} and only more recently for
what concerns mechanical folding and refolding \cite{cinpull,lacks}.
The model consists of a chain of $L$ point-like
monomers mimicking the residues of a polypeptidic chain. For the sake of
simplicity, only three types of residues are considered: hydrophobic (B),
polar (P) and neutral (N) ones. 

The intramolecular potential is composed of four terms: a stiff
nearest-neighbour harmonic potential, $V_1$, intended to maintain the bond
distance almost constant, a three-body interaction $V_2$, which accounts for the
energy associated to bond angles, a four-body interaction $V_3$ corresponding to
the dihedral angle potential, and a long--range Lennard-Jones (LJ) interaction, $V_4$,
acting on all pairs $i$, $j$ such that $|i-j| >2$, namely
\begin{eqnarray}
V_1 (r_{i,i+1}) &=& \alpha (r_{i,{i+1}}-r_0)^2, 
\label{harm}\\
V_2(\theta_i) &=& A \cos(\theta_i) +B \cos(2 \theta_i) - V_0,
\label{bond}\\
V_3(\varphi_i, \theta_i, \theta_{i+1}) &=& C_i [1-S(\theta_i)S(\theta_{i+1})\cos(\varphi_i))]
 + D_i [1-S(\theta_i)S(\theta_{i+1})\cos(3 \varphi_i))], 
\label{dih}\\
V_4(r_{i,j}) &=& \varepsilon_{i,j} \left( \frac{1}{r_{i,j}^{12}} - \frac{c_{i,j}}{r_{i,j}^6} \right)
\label{lj}
\end{eqnarray}
Here, $r_{i,j}$ is the distance between the $i$-th and the $j$-th monomer,
$\theta_i$ and $\varphi_i$ are the bond and dihedral angles at the $i$-th monomer, respectively.
The parameters $\alpha =50$ and $r_0 =1$ (both expressed in adimensional units) fix
the strength of the harmonic force and the equilibrium distance between
subsequent monomers (which, in real proteins, is of the order of a few \AA).
The value of $\alpha$ is chosen to ensure a value for $V_1$
much larger than the other terms of potential in order
to reproduce the stiffness of the
protein backbone. The expression for the bond-angle potential term $V_2 (\theta_i)$ (\ref{bond}) corresponds, 
up to the second order, to a harmonic interaction term $\sim (\theta_i -\theta_0)^2/2$, where
\begin{equation}
A=- k_{\theta} \frac{cos(\theta_0)}{\sin^2(\theta_0)} ,\qquad
B= \frac{k_{\theta}}{ 4 \sin^2(\theta_0)} ,\qquad
V_0= A \cos(\theta_0) + B \cos( 2\theta_0) \quad ,
\label{param_harm}
\end{equation}
with $k_{\theta} = 20 \epsilon_h$, $\theta_0 = 5 \pi/12 \enskip rad$ or $75^o$ and
where $\epsilon_h$ sets the energy scale.
This formulation in terms of cosines allows to speed up the simulation, since it is sufficient 
to evaluate $\cos(\theta_i)$ and the value of bond-angle is not needed, and at the same time to 
avoid spurious divergences in the force expression due to the vanishing of $\sin(\theta_i)$ 
when three consecutive atoms become aligned \cite{karplus}.

\begin{figure}[t]
\includegraphics[draft=false,clip=true,height=0.34\textwidth]{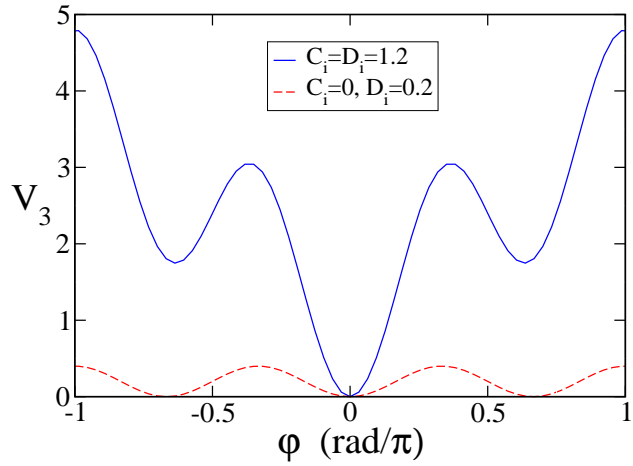}
\caption{(Color online) Dihedral angle potential, $V_{3}$, when two or more beads among the four defining
$\varphi$ are neutral (red dashed curve), and in all the other cases (blue solid curve). 
We fixed $\varepsilon_h = 1$ and $S(\theta_i)=S(\theta_{i+1}\equiv 0$.
}
\label{pot_dih}
\end{figure}

The dihedral angle potential is characterized by three minima for $\varphi = 0$ (associated to
a so-called {\it trans state})  and $\varphi= \pm 2 \pi/3$ (corresponding to {\it gauche states}),
this potential is mainly responsible for the formation of secondary structures. In particular
large values of the parameters $C_i, D_i$ favor the formation of trans state and therefore
of $\beta$-sheets, while when gauche states prevail $\alpha$-helices are formed. The parameters
$(C_i,D_i)$ have been chosen as in \cite{veit}, i.e. if two or more beads among the four defining
$\varphi$ are neutral (N) then $C_i = 0$ and $D_i = 0.2 \varepsilon_h$, in all the other cases
$C_i = D_i = 1.2 \varepsilon_h$ (see Fig.~\ref{pot_dih}). The {\it tapering function} $S(\theta_i) = 1 - \cos^{32} (\theta_i)$ has been
introduced in the expression of $V_3$ in order to cure a well known problem in the dihedral 
potentials \cite{karplus}. This problem is encountered whenever $\theta_i = 0$ or $\pi$, i.e. when three consecutive
beads are in the same line, in these situations the associated dihedral angle is no more defined and 
a discontinuity in $V_3$ arises. In contrast to what reported in \cite{karplus} this situation is 
not improbable for the present model. The quantity $S(\theta_i)S(\theta_{i+1})$ entering in the definition 
of $V_3$ has a limited influence on the dynamics apart in proximity of the above mentioned extreme cases.
Moreover, $S(\theta_i)S(\theta_{i+1})$ is $C^\infty$,
its value is essentially one almost for any $\theta_i$, it does not introduce 
any extra minima in the potential and it vanishes smoothly for $\theta_i \to 0$ or $\theta_i \to \pi$
\cite{rampioni}.

The last term $V_4$ has been introduced to mimic effectively the interactions with the solvent,
it is a Lennard-Jones potential and it depends on the type of interacting residues as follows:
\begin{itemize}

\item if any of the two monomers is neutral the potential is repulsive $c_{N,X} =0$ 
and its scale of energy is fixed by $\varepsilon_{N,X} = 4 \varepsilon_h$;

\item for interactions between hydrophobic residues $c_{B,B} =1$ and $\varepsilon_{B,B} = 4 \varepsilon_h$;

\item for any polar-polar or polar-hydrophobic interaction $c_{P,P} \equiv c_{P,B} = -1$ 
and $\varepsilon_{P,P} \equiv \varepsilon_{P,B} = (8/3) \varepsilon_h$.

\end{itemize}

Accordingly, the Hamiltonian of the system reads
\begin{eqnarray}
H = K + V = \sum_{i=1}^L \frac{p_{x,i}^2+p_{y,i}^2+p_{z,i}^2}{2} +\sum_{i=1}^{L-1} V_1(r_{i,i+1}) +
\nonumber\\
+\sum_{i=2}^{L-1} V_2(\theta_i)+ 
\sum_{i=2}^{L-2} V_3(\varphi_i,\theta_i,\theta_{i+1}) +
\sum_{i=1}^{L-3} \sum_{j=i+3}^{L}  V_4(r_{ij})
\label{hamil}
\end{eqnarray}
where, for the sake of simplicity, all monomers are assumed to have the same
unitary mass, the momenta are defined as $(p_{x,i},p_{y,i},p_{z,i}) 
\equiv ({\dot x}_i,{\dot y}_i,{\dot z}_i)$
and we fix $\varepsilon_h = 1$.

In the present paper we consider the two following sequences of 46 monomers,

\begin{itemize}

\item{[GF]=$B_9N_3(PB)_4  N_3 B_9  N_3 (PB)_5P$} a sequence that has been
widely analyzed in the past for spontaneous folding \cite{honey,guo_thir,guo,veit,berry,evans,kim,kim-keyes} as well
as for mechanical unfolding and refolding \cite{cinpull,lacks};

\item{[BF]= $BNBPB_{3}NPB_{4}NBPB_{2}NP_{2}B_{5}N_{2}BPBNPB_{2}NBP_{2}BNB_{2}PB_{2}$ } a randomly 
generated sequence, but with the same number of B, P and N monomers as the $GF$.
\end{itemize}

These two sequences have been chosen because $GF$ has been previously identified as
a reasonably fast folder \cite{guo_thir} (see also \cite{berry} for a detailed and critical 
analysis of the basin-bottom structures observed for this model), while we expect that the sequence $BF$, being randomly 
chosen, cannot have the characteristic of a good folder. 
From now on we refer to the sequence $GF$ (resp. $BF$) as the {\it good} (resp. {\it bad}) folder.

The $46$-mer sequence $GF$ exhibits a four stranded $\beta$-barrel
Native Configuration (NC) with an associated potential energy $E_{NC}=-49.878$. Please
note that the model is here analyzed by employing the same potential and parameter set 
reported in Ref. \cite{veit}, but neglecting any diversity among the hydrophobic residues. 
The NC, displayed in Fig. \ref{config.gf}(a),
is stabilized by the attractive hydrophobic interactions among the $B$ residues,
in particular the first and third $B_9$ strands, forming the core of the NC,
are parallel to each other and anti-parallel to the second and fourth strand,
namely, $(PB)_4$ and $(PB)_5P$. The latter strands are exposed towards the
exterior due to the presence of polar residues.

As shown in  Fig. \ref{config.bf}, the native structure of the $BF$ is quite different, 
it has a core constituted by the first three $\beta$-strands and a very long "tail" (made of 18 residues)
wrapped around the core. In particular, the first and second $\beta$-strand 
(namely, $BNBPB_{3}NP$ and $B_{4}NBPB_{2}$) are formed by 9 residues, and 
anti-parallel to each other. For more clarity, we will term $\pi_1$
the plane containing the first 2 strands.
The third strand 
(namely, $P_{2}B_{5}$) is made of 7 residues and it is 
located in a plane lying in-between the first and second strand, which is
almost perpendicular to $\pi_1$.
The chain rotates of almost $90$ degrees in correspondence of the two consecutive
neutral beads and then exhibits a short strand of 3 beads $PBP$ before
turning back with a parallel strand of 7 beads ($PB_{2}NBP_{2}$) that
passes below $\pi_1$. Finally the
chain turns once more back by passing this time above the plane $\pi_1$.
In the final part of the tail of the chain a short strand of
5 residues, parallel to the $4$-th and $5$-th strands, can be identified 
as $B_{2}PB_{2}$. The potential energy of the NC of the $BF$ is quite 
high with respect to the $GF$, namely $V_{NC} = -23.956$. Moreover,
this difference, as reported in Table \ref{NC}, is essentially due
to the difference in the dihedral contributions, that is much higher in the NC of the
$BF$ with respect to the $GF$, while  all the other contributions,
in particular the LJ ones, have nearby values.
The dihedral contribution that arises in the $BF$ is essentially 
due to the configuration of the first 3 strands, since these
are arranged over two almost orthogonal planes.

\begin{table}
\begin{center}
\begin{tabular}{r|r|r} \hline\hline 
\hspace{0.3cm}   & \hspace{0.7cm} GF \hspace{0.2cm}  & \hspace{0.2cm} BF \hspace{0.2cm} \\ \hline
$V_{NC}$     &  -49.878           & -23.956    \\ 
$V_1$ \hfill       &    0.787           & 0.777    \\ 
$V_2$  \hfill      &    1.767           & 5.744    \\ 
$V_3$   \hfill     &    2.602           & 23.105    \\ 
$V_4$   \hfill     &   -55.035          & -53.582    \\ \hline \hline
\end{tabular}
\end{center}
\caption{Potential energy values associated to the NC of the GF and BF, the different
contributions to the total potential energy $V_{NC}$ are also reported.
}
\label{NC}
\end{table}

\subsection{Simulation protocol: equilibrium Langevin dynamics}
\label{simulation_protocol}

Molecular dynamics (MD) canonical simulations at equilibrium temperature $T$
have been performed by integrating the corresponding 
Langevin equation for each monomer of unitary mass (characterized by the position vector ${\bf r}_{i})$):
\begin{equation}
\ddot{{\bf r}}_{i}  = {\bf F}({\bf r}_{i}) - \gamma\dot{{\bf r}}_{i} + {\bf \eta}(t)  \hspace{1.0cm} i=1,L 
\end{equation}
where  ${\bf \eta}(t)$ is a zero average Gaussian noise term with correlations given by
$\langle \eta_{\alpha}(t) \eta_{\beta}(t ^{\prime}) \rangle = 2 T \gamma \delta (t-t ^{\prime})
\delta_{\alpha,\beta}$; ${\bf F}=-{\bf \nabla} V$, being $V$ the intramolecular potential introduced in \ref{model},
$\gamma$ the friction coefficient associated to the solvent and by assuming an unitary Boltzmann
constant.

Numerical integrations have been implemented via a standard 
Euler scheme with a time-step $\Delta t=0.005$ and with a low friction
coefficient $\gamma = 0.05$ \cite{veit}. Two different kinds of MD have been performed, namely unfolding 
simulations (US) and folding simulations (FS). In the first case the initial state
of the system is taken equal to the native configuration (NC), that we assume 
to coincide with the minimal energy configuration.
In the latter one the initial state is a completely unfolded configuration.
 
\subsection{Simulation protocol: out-of-equilibrium mechanical unfolding}

 In order to mimic the mechanical pulling of the protein attached to a AFM cantilever,
or analogously when trapped in an optical tweezer, one extremum of the chain was kept fixed and the last bead
is attached to a pulling apparatus with a spring of elastic constant $k$. The external
force is applied by moving the "cantilever" along a fixed direction with a certain protocol
$z(t)$. Before pulling the protein, the coordinate system is always rigidly
rotated, in order to have the z-axis aligned along the end-to-end direction
connecting the first and last bead.  Therefore by denoting with
$\zeta$ the end-to-end distance the component of the external force along 
this direction reads as
\begin{equation}
F_{ext} = k (z - \zeta)
\end{equation}
where $k=10$ in order to suppress fast oscillations.
As recently pointed out it is extremely important to use a sample of thermally 
equilibrated initial configurations to correctly reproduce the equilibrium FEL via the JE.
\cite{west_paci}.
Therefore, before pulling the protein, we have performed a thermalization
procedure in two steps. At a fixed temperature $T$,
initially the protein evolves freely starting from the NC for a time
$t=1,000$, then it is attached
to the external apparatus, with the first bead blocked, and it
equilibrates for a further for a further time period $t = 500$. 
The system (at sufficiently low temperatures) quickly settles down to a "native-like" configuration. 
This configuration is then employed as the starting state for the forced folding. 
The protocol that we have 
used is a linear pulling protocol with a constant speed $v_p$, i.e. $z(t) = z(0) + v_p \times t$,
by assuming that the pulling starts at $t=0$.  Usually we have employed velocities
$ v_P \in [ 5 \times 10^{-6}: 5 \times 10^{-2}]$ and set $z(0) = \zeta_0$, i.e. to 
the end-to-end distance associated to the native configuration.

\section{Weighted histogram analysis method}\label{three_I}

 A combination of the umbrella sampling technique \cite{umbrella} 
with the weighted histogram analysis method (WHAM) \cite{wham}
allows to obtain the equilibrium
free energy profile as a function of the end-to-end distance.

The umbrella sampling technique \cite{umbrella} amounts  to perform of a series of biased molecular 
dynamics simulations of the system constrained by  an external potential, namely
\begin{equation}
w_i(\zeta) = \frac{1}{2} k_W (\zeta - {\bar \zeta}_i)^2 \qquad.
\label{wham_pot} 
\end{equation}
The potential $w_i$ forces the heteropolymer to stay in configurations
characterized by a certain average
end-to-end distance ${\bar \zeta}_i$, even if at the considered temperature
such $\zeta$-value is highly unfavored. These simulations allows
to obtain a series of $M$ biased end-to-end probability density 
distributions $\rho_i^B(\zeta) \{i=1,\dots, M\}$, which properly combined
can permit the reconstruction of the equilibrium unbiased $\rho(\zeta)$.
In particular, in the case of identical statistics for each biased run
the WHAM formalism prescribes the following combination
\begin{equation}
\rho(\zeta) = \frac{\sum_{i=1}^M \rho_i^B(\zeta)}{\sum_{i=1}^M {\rm e}^{-\beta[w_i(\zeta)-F_i]}}
= {\rm e}^{-\beta f_W(\zeta,T)}
\label{wham} 
\end{equation}
where $\beta=1/T$ and the free energy constants $\{F_i\}$ can be obtained by the normalization condition
\begin{equation}
{\rm e}^{-\beta F_i} = \int d\zeta \enskip {\rm e}^{-\beta w_i(\zeta)} \rho(\zeta)
\qquad .
\label{FFF} 
\end{equation}
Eqs. (\ref{wham}) and (\ref{FFF}) should be solved self-consistently 
via an iterative procedure, finally this allows to obtain an estimate
of the equilibrium free energy $f_W(\zeta,T)$, apart from an additive constant.

We have considered equally spaced $\{{\bar \zeta}_i\}$-values, with a separation
$\Delta {\bar \zeta}_i = 0.2$ among them, ranging from the native configuration
$\zeta_0$ to the all {\it trans}-configuration $\zeta_{trans}$
\footnote{This is an elongated (planar) equilibrium conformation of the protein with all 
the dihedral angles at their {\it trans} values, corresponding to $\zeta_{trans}=35.70$.}
. For each of the $M$ runs,
after a quite long equilibration time $t \sim 120,000 - 200,000$, we have estimated 
$\rho_i^B(\zeta)$ over 100,000 configurations taken at regular time intervals $\Delta t= 0.2$.
The biased simulations have been performed with a hard and weak spring, corresponding to
$k_W=10$ and 0.5 in (\ref{wham_pot}), respectively. The results obtained
essentially agree for the two $k_W$-values, apart when the free energy landscape
exhibits steep increases as a function of $\zeta$. In these cases the
hard spring is more appropriate, since the weak one allows the protein to 
refold, thus rendering the $\zeta$-intervals, where $f_W(\zeta)$ is steeper,
not accessible to the WHAM reconstruction.

\section{Inherent structure formalism}\label{three}

 Inherent structures correspond to local minima of the potential energy, 
in particular the phase space visited by the protein during its
dynamical evolution can be decomposed into disjoint attraction basin, each corresponding to 
a specific IS. Therefore, the  canonical partition function can be expressed within the IS formalism as 
a sum over the non overlapping basins of attraction, each corresponding to a specific minimum (IS) $a$ 
\cite{wales,nakagawa}: 
\begin{equation}
Z_{IS}(T) = 
\frac{1}{\lambda^{3N^\prime}} 
\sum_a {\rm e}^{-\beta V_a} \int_{\Gamma_a}
{\rm e}^{-\beta \Delta V_a (\Gamma)} d \Gamma  =
\sum_a {\rm e}^{-\beta  [V_a+R_a(T)]} 
\label{zeta} 
\end{equation}
where $N^\prime$ is the number of degrees of freedom of the system, 
$\lambda$ is the thermal wavelength, $\Gamma$ represents one of the possible conformations of 
the protein within the basin of attraction of $a$,
$V_a$ is the potential energy associated to the minimum $a$, $\Delta V_a (\Gamma)=V(\Gamma)-V_a$ and
$R_a(T)$ the vibrational free energy due to the fluctuations around the minimum.

The vibrational term $R_a(T)$ can be estimated by assuming a harmonic basin of attraction:
\begin{equation}
{\rm e}^{-\beta R_a(T)} = 
\frac{1}{\lambda^{3N-6}} 
\int_{\Gamma_a} {\rm e}^{-\beta \Delta V_a (\Gamma)} d \Gamma
= \prod_{j=1}^{3N-6} \frac{T}{\omega_a^j}
\label{fvib} 
\end{equation}
where $\omega_a^j$ are the frequencies of the vibrational modes around the IS $a$
and an unitary reduced Planck constant has been considered.

Therefore the probability  to be in the basin of attraction of the IS $a$ is
\begin{equation}
p_a(T)= \frac{1}{Z_{IS}(T)} {\rm e}^{-\beta (V_a + R_a(T))}  \qquad .
\label{pa} 
\end{equation}
 
The free energy of the whole system at equilibrium is simply given by
$ f_{IS}(T)= - T \ln[Z_{IS}(T)]$. However if one is interested to construct
a free energy landscape as a function of a parameter characterizing the
different IS, like e.g. the Kabsch distance $\delta_K$ \cite{kabsch} or the end-to-end
distance $\zeta$, this is possible by defining a partition function 
restricted to IS with an end-to-end distance within the narrow interval
$[\zeta; \zeta +d \zeta]$
\begin{equation}
Z_{IS}(\zeta, T) = 
{\sum_a}^\prime {\rm e}^{-\beta [V_a+R_a(T)]} 
\label{zetaprime} 
\end{equation}
where the $\sum^\prime$ indicates that the sum is not over the whole ensemble 
of ISs $\{ a \}$ but restricted.  
The free energy profile as a function of $\zeta$ can be simply obtained by the
relationship:
\begin{equation}
f_{IS}(\zeta,T)= - T \ln[Z_{IS}(\zeta, T)] \qquad ;
\label{FIS} 
\end{equation}
while the average potential and free vibrational energy, corresponding to ISs 
characterized by a certain $\zeta$, can be estimated as follows:
\begin{equation}
V_{IS}(\zeta,T) = \frac{{\sum_a}^{\prime} V_a \enskip {\rm e}^{-\beta [V_a+R_a(T)]}}{Z_{IS}(\zeta,T)} \qquad ;
\qquad R_{IS}(\zeta,T) = \frac{{\sum_a}^{\prime} R_a(T) \enskip {\rm e}^{-\beta [V_a+R_a(T)]}}{Z_{IS}(\zeta,T)} \qquad .
\label{VIS_RIS} 
\end{equation}

In order to find the different ISs
one can perform MC samplings or molecular dynamics (MD) simulations. We have
chosen to examine MD trajectories at constant temperature via a Langevin 
integration scheme. In particular, we have built up two data banks of ISs:
the thermal data bank (TDB) obtained by performing equilibrium canonical simulations and the 
pulling data bank (PDB) by mechanically unfolding the protein. 
In order to find the different ISs the equilibrium 
(resp. out-of-equilibrium) Langevin trajectory is sampled at constant time intervals 
$\delta t = 5$ (resp. at constant elongation increments 
$\delta \zeta = 0.1$) to pinpoint a series 
of configurations, which afterward are relaxed via a steepest descent dynamics
and finally refined by means of a standard Newton's method. 
In the case of the TDB, in order to speed up the search of ISs we have employed a 
so-called "quasi-Newton" method \cite{numerical} \footnote{The comparison between the
steepest descent and the quasi-Newton methods has revealed that this second minimization scheme
is somehow faster (1.8 times faster at $T=0.5$ for the good folder), but while the steepest
descent algorithm
is able to identify the metastable stationary states in the 99.8 \% of examined cases the 
quasi-Newton scheme
was successful in the 98.7 \% of situations. However the distributions of the identified
minima (by considering the same trajectory) obtained with the two schemes are essentially coincident.}.
For mechanical unfolding, the protein is unblocked and the pulling apparatus removed
before the relaxation stage. Two local minima are identified as distinct whenever their
energies differ more than $1 \times 10^{-5}$.
The TDB for the good (resp. bad) folder contains $579,749$ (resp. $210,782$) distinct ISs 
collected via equilibrium simulations at various temperatures in the range $[0.3;2.0]$. The PDB
contains $3,000 - 50,000$ ISs depending on the examined temperature as detailed in the Table
\ref{IS_PDB}.

\begin{table}
\begin{center}
\begin{tabular}{r|r|r} \hline\hline 
\hspace{0.3cm} T  & \hspace{0.7cm} goodfolder \hspace{0.7cm}  & \hspace{0.7cm} badfolder \hspace{0.7cm} \\ \hline
$0.1$        & 2,843                               & 456    \\ 
$0.2$        & 5,875                               & 1,763   \\ 
$0.3$        & 12,359                              & 6,477   \\ 
$0.4$        & 35,409                              & 21,060    \\
$0.5$        & 52,546                              & 45,950    \\ 
$0.6$        & 51,971                              & ---    \\ 
$0.7$        & 54,736                              & ---    \\ \hline \hline
\end{tabular}
\end{center}
\caption{Number of distinct ISs contained in the PBD at different temperatures.
These have been obtained by sampling, during out-of-equilibrium
mechanical unfoldings, several Langevin trajectories at constant elongation 
increments $\delta \zeta = 0.1$. The total number of relaxations performed
for each temperature amounts to $\sim 60,000$ corresponding to $\sim 200$
repetitions of the same pulling experiment. The considered experiments
have been performed at $v_P = 5 \times 10^{-4}$ for the GF,
while velocities in the range $ v_P \in [ 5 \times 10^{-5}: 5 \times 10^{-4}]$ 
have been employed for the BF.
For the bad folder not all 
temperatures have been examined.
}
\label{IS_PDB}
\end{table}

\section{Extended Jarzynski equality}\label{four}

In the present section, we discuss an extended version of the Jarzynski equality, 
which allows one to obtain the free energy profile as a function of a collective coordinate.
Let $x$ be the variable that  identifies the system microscopic state, e.g. 
the collection of the positions and momenta of all the particles in the system $x=\{\bm r_i,\bm p_i\}$. 
The system Hamiltonian is a function of $x$, and will be indicated as $H_0(x)$ in the following.
Let $X(x)$ be a macroscopic observable of the system, e.g. the volume, 
and let us assume that the system is subject to an external potential $U_\lambda(X)$, which is function of $X$, and which depends on a parameter $\lambda$ whose value is externally controlled. The parameter $\lambda$ changes according to a given time protocol $\lambda(t)$, and thus the system is characterized by a time dependent Hamiltonian
$H(x,t)=H_0(x)+U_{\lambda(t)}(X(x))$. The thermodynamic work done on the system, as the external parameter $\lambda$ changes, reads
\begin{equation}
W_t=\int_0^t \D t'\;\dot \lambda(t')
\left.\partial_\lambda U_\lambda(X(x(t'))\right|_{\lambda=\lambda(t')}.
\label{work_def}
\end{equation}
Due to thermal fluctuations, $W_t$ varies between a realization 
and another one of the manipulation process.

We now introduce the function $f(X,T)$, which is the free energy of the constrained ensemble, in which the value $X(x)$ is fixed at $X$:
\begin{equation}
f(X,T)=- k_B T \ln\int \D x\;\delta(X-X(x))\,\E^{-\beta H_0(x)} .
\label{f_X}
\end{equation}

The extended Jarzynski equality, relates the work done on the system, as an effect of
the change in the external parameter $\lambda$, with the free energy $f(X,T)$ ~\cite{HumSza,seifprl,imparato}:
\begin{equation}
Z_0  \E^{\beta U_{\lambda(t)}(X)}\average{\delta(X-X(x))\E^{-\beta W_t}}_t
  =\E^{-\beta f(X,T)},
\label{sample}
\end{equation}
where $Z_0=\int \D x \exp\left[-\beta H_0(x)\right]$ is the partition function associated with the 
time-independent Hamiltonian $H_0(x)$ and the averages $\average{\cdot}_t$ are taken over
many realizations of the same protocol at time $t$.  Equation~(\ref{sample}) provides thus a method to evaluate the
unperturbed free energy $f(X,T)$ as long as one has a reliable
estimate of the lhs of this equation.
It is worth to note that one does not need to evaluate the partition function $Z_0$ to evaluate $f(X,T)$, as it appears only as a multiplicative constant in eq.~(\ref{sample}).

The optimal estimate of $f(X,T)$ can be obtained by combining Eq.~(\ref{sample}) 
with the previously discussed method of weighted
histograms ~\cite{HumSza,seifprl,imparato}, namely
\begin{equation}
f_J (X,T)=- k_B T\ln \pq{\frac{\sum_t\frac{\average{\delta(X-X(x)) 
\exp\p{-\beta W_t}}_t}{\average{ \exp\p{-\beta W_t}}_t}}{\sum_t\frac{\exp\p{-U(X,t)}}{ \average{\exp\p{-\beta W_t}}_t}}}
\quad ;
\label{ext_jarz}
\end{equation} 
where the sums $\sum_t$ are over successive time snapshots. For a detailed derivation of 
Eq.~(\ref{sample}) see ~\cite{imparato}.

\section{Thermodynamical properties}
\label{five}

\begin{figure}[t]
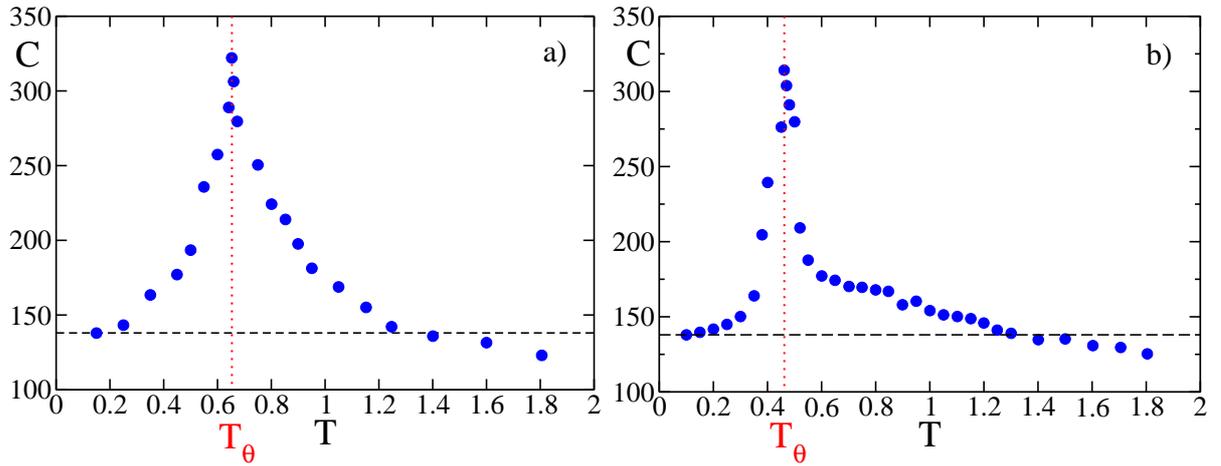

\includegraphics[draft=false,clip=true,height=0.34\textwidth]{f2a}
\includegraphics[draft=false,clip=true,height=0.34\textwidth]{f2b}
\caption{(Color online) Heat capacity $C$ as a function of the temperature $T$ for good (a) and bad (b) folder; 
the vertical (red) dotted line indicates the hydrophobic collapse temperature 
$T_\theta$ and the horizontal (black) dashed line
the value $C_{sol}$.}
\label{Ttheta}
\end{figure}

The main thermodynamical features of the examined model can be summarized by reporting three different transition 
temperatures \cite{wales,baum,tlp,tap,kim-keyes}: namely, the hydrophobic collapse temperature $T_\theta$, the folding 
temperature $T_f$, and the glassy temperature $T_g$.
\\
The collapse temperature discriminates between phases
dominated by random-coil configurations rather than collapsed ones \cite{degennes}, 
$T_\theta$  has been usually identified as the temperature where the heat capacity $C(T)$ reaches its maximal value, namely
(within the canonical formalism):
\begin{equation}
C(T_\theta) \equiv C^{max} \quad, 
\hspace{0.5cm} {\rm where} \hspace{0.5cm} C(T)=\frac{\langle E^2 \rangle-\langle E\rangle^2}{T^2},
\end{equation}
and $<\cdot >$ represents a time average performed over an interval $t \simeq 10^5$ by following an US trajectory.
From Fig. \ref{Ttheta}, it is evident that for both sequences $C(T) \sim 138$ up to temperatures
$T \sim 0.25$. This result can be understood by
noticing that at low temperatures the thermal features of heteropolymers resemble that of a 
disordered 3D solid, with an associated heat capacity $C_{sol} \equiv 3L$. Moreover, 
the high temperature values are smaller than $C_{sol}$, since in this limit we expect that 
a one dimensional chain in a three dimensional
space would have a specific heat $C=2 L$ \cite{tlp}. 
However, as shown in Fig. \ref{Ttheta}, these extreme temperatures have not yet been reached.
The comparison of the heat capacity curves for the GF and BF
reveals that $C(T)$ obtained for the GF has a much broader peak with respect
to the BF.  This indicates that the transition from the
NC to the random coil state is definitely sharper for the bad folder.

\begin{figure}[t]
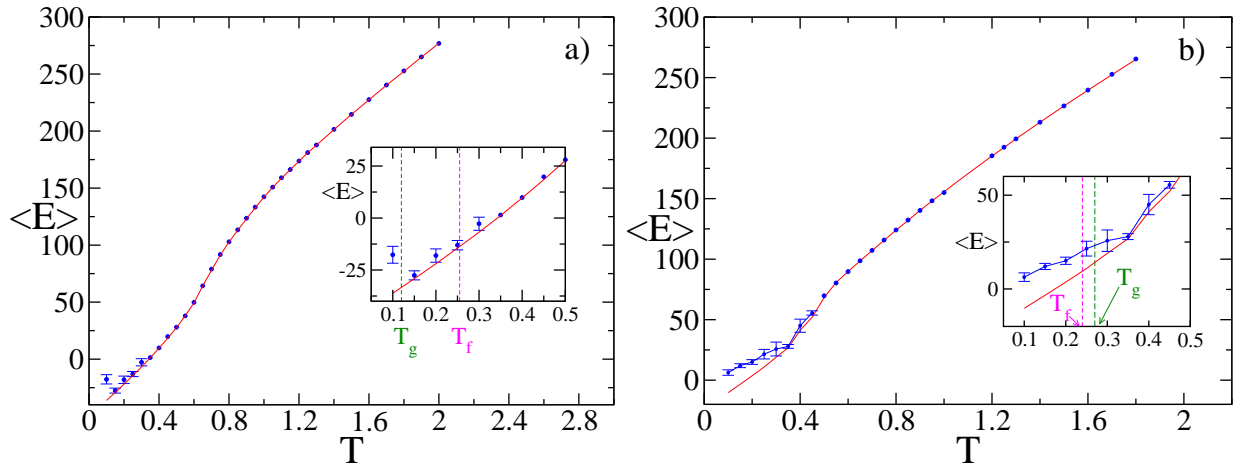

\includegraphics[draft=false,clip=true,height=0.34\textwidth]{f4a.eps}
\includegraphics[draft=false,clip=true,height=0.34\textwidth]{f4b.eps}
\caption{(Color online) Total energy $\langle E \rangle$ as a function of the temperature $T$ for good (a) and bad (b) folder; 
the solid (red) line corresponds to US's and the
 (blue) symbols to FS's. In the inset an enlargement for low temperatures: the dashed lines indicate the glassy ($T_g$) (magenta)
  and folding ($T_f$) (green) temperatures.}
\label{Tglassy}
\end{figure}

The folding temperature has been defined in many different ways \cite{guo,veit,tlp}, however
we have chosen to define the folding temperature by employing the IS reconstruction of the phase space.
In practice, quite long USs have been performed at various temperatures ,
up to duration $t= 5,000,000$. During each of this US the visited ISs have been identified
at regular intervals $\delta t = 5$, and from these data we have estimated the probability
$P_{nc}(T)$ to visit the NC at such temperature. 
The folding temperature $T_f$ is then defined as
\begin{equation}
P_{nc}(T_f) \equiv 0.5  \qquad .
\end{equation}
Indeed, it should be noticed that for the GF $P_{nc}$ is the probability
to stay in the two lowest lying energy minima (ISs) and not in the NC only.
These two minima can be associated to an unique attraction basin,
since their energy separation is extremely small with respect to $|V_{NC}|$ 
(namely, $0.04$) and also the corresponding configurations are almost
identical, being separated by a Kabsch distance $\delta_K = 0.128$.
Moreover, at any examined temperature we have always observed a rapid switching between the
two configurations, indicating that there is an extremely low energy barrier among these
two states.
\\
\begin{figure}[t]
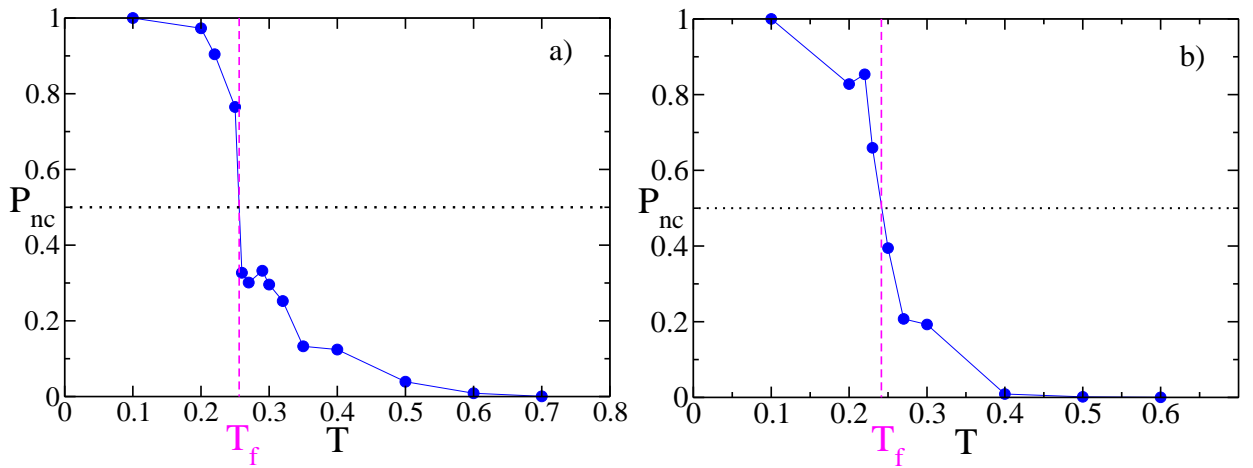

\includegraphics[draft=false,clip=true,height=0.34\textwidth]{f3a.eps}
\includegraphics[draft=false,clip=true,height=0.34\textwidth]{f3b.eps}
\caption{(Color online) Probability $P_{nc}$ as a function of the temperature $T$ for good (a) and bad (b) folder; 
the vertical (magenta) dashed line indicates the folding temperature $T_f$, while the horizontal (black) dotted line 
refers to the value $0.5$.
.}
\label{Tfolding}
\end{figure}
The glassy temperature $T_g$ indicates the temperature below which freezing of 
large conformational rearrangements occurs:
below such a temperature the system can be trapped in local minima of the potential. 
By following \cite{tlp}, in order to locate $T_g$ we have made a comparison 
among results obtained from FS and US. In particular, we have examined, at the same temperatures,
the average total energy $\langle E \rangle$ of the system evaluated over finite time intervals.
As shown, in Fig. \ref{Tglassy}, these quantities, when obtained from USs and FSs,
coincide at temperatures larger than $T_g$, below which the structural arrest takes place.
In particular, unfolding averages have been performed over intervals of duration $t = 10^5$ by following a 
single trajectory. On the other hand, folding simulations have been followed up to
times $t \simeq 1.1 \cdot 10^7$ and the averages taken over 5-7 different initial 
conditions by considering for each trajectory only the last time span of duration $t \simeq 5 \cdot 10^4$.
The error bars (standard deviation) shown in Fig. \ref{Tglassy} should be interpreted , at
sufficiently low temperatures, as a sign of the dependence of the results on the initial conditions. 
\\
The three transition temperatures estimated for the good and bad folder are reported in table \ref{transition_temperatures} 
\footnote{In \cite{guo} for the sequence $GF$ it has been found $T_\theta = 0.65$ and $T_f \sim 0.34$; moreover in the same paper the authors
suggested that the folding transition was associated to a shoulder in the $C$, but this result has been recently criticized \cite{kim}.
Moreover, more recent estimates, obtained by employing different protocols,  
suggest that $T_f \sim 0.24-0.25$ \cite{evans,kim-keyes}
and $T_g \sim 0.15 $ \cite{kim-keyes}, values that are essentially in agreement with our results 
}.
\begin{table}
\begin{center}
\begin{tabular}{r|r|r} \hline\hline 
  \hspace{0.3cm}  & \hspace{0.7cm} GF \hspace{0.7cm}  & \hspace{0.7cm} BF \hspace{0.7cm} \\ \hline
$T_\theta$        & 0.65(1)                                   & 0.46(2)    \\ 
$T_f$             & 0.255(5)                                  & 0.24(1)    \\
$T_g$             & 0.12(2)                                   & 0.27(2)    \\ \hline \hline
\end{tabular}
\end{center}
\caption{Transition temperatures estimated for good and bad folder with the corresponding error.}
\label{transition_temperatures}
\end{table}
One can notice that $T_\theta$ is larger for the good folder, thus indicating that the collapsed state
has a greater stability with respect to the bad folder. Moreover, while for the good folder $T_f > T_g$,
for the bad one this order is reversed. Therefore the BF will most likely remain trapped
in some misfolded configurations before reaching the NC even at temperatures $T \sim T_f$.

\section{Extended Jarzynski equality reconstruction}
\label{six}

In this section we present for both the sequences $GF$ and $BF$ the reconstruction of the FEL,
at various temperatures, as a function of the end-to-end distance $\zeta$ starting
from out-of-equilibrium measurements.
The free energy profiles have been obtained via the EJE by averaging over $28-250$ repetitions 
of the same pulling protocol depending on the pulling velocity 
as described in section \ref{simulation_protocol}.
We have generally used the pulling configuration where the first bead is kept fixed 
and the 46th bead is pulled (tail-pulled case). 
However, by considering the head-pulled case, 
where the roles of the first and last bead are reversed.
we obtain, for sufficiently low velocities (namely, $ v_p \le 5 \times 10^{-4}$ for the GF 
and $ v_p \le 5 \times 10^{-5}$ for the BF), 
exactly the same free energy profile. These results are essentially in
agreement with those reported in \cite{cinpull} for the GF.

\subsection{Good folder}

\begin{figure}[t]
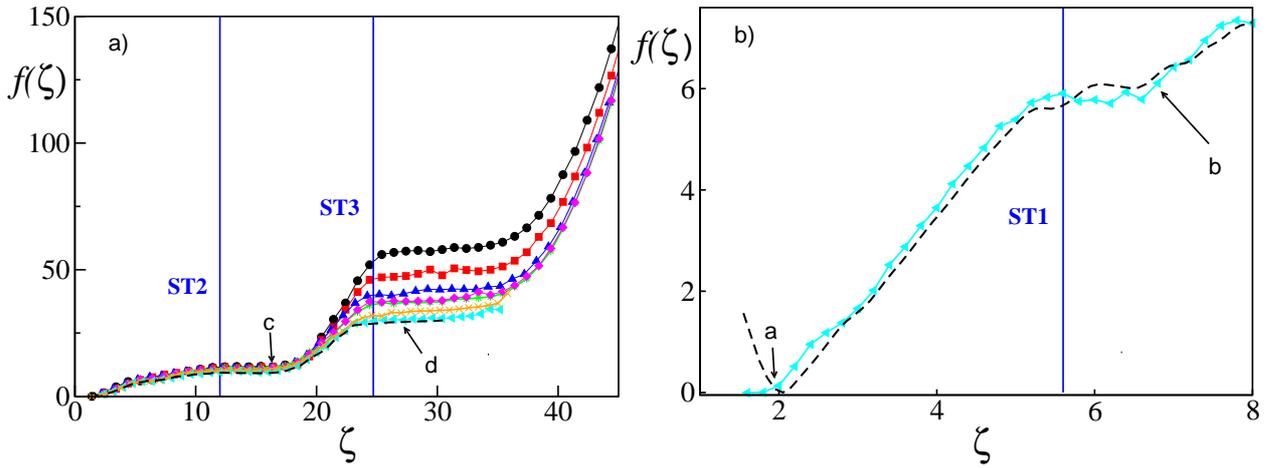

\includegraphics[draft=false,clip=true,height=0.34\textwidth]{f5a}
\includegraphics[draft=false,clip=true,height=0.34\textwidth]{f5b}
\caption{(Color online) (a) Free energy profiles $f_J$ for the good folder as 
a function of the end-to-end distance $\zeta$ 
at $T=0.3$, obtained with the EJE for various pulling velocities: from top to bottom
$v_p=5 \times 10^{-2}$, $1 \times 10^{-2}$, $5 \times 10^{-3}$,  $5 \times 10^{-4}$,
$2 \times 10^{-4}$, $2 \times 10^{-5}$ and $5 \times 10^{-6}$. 
In (b)  an enlargement of the curve for $v_p=5 \times 10^{-6}$ at low $\zeta$ is reported. 
The (black) dashed curve in (a) and (b) 
refers to the WHAM reconstruction $f_W(\zeta)$ with $k_W=10$.
The number of different pulling experiments performed to estimate the profiles
ranges between $150$ and $250$ at the higher velocities
to $28$ at the lowest velocity $v_p=5 \times 10^{-6}$.
The letters indicate the value of $f(\zeta)$ for the 
pulled configurations reported in Fig. \ref{config.gf} (a)
and the (blue) vertical solid lines the location of the STs.}
\label{gf_eje_t0.3}
\end{figure}

In Fig.~\ref{gf_eje_t0.3} (a) are presented the EJE reconstructions $f_J(\zeta)$ 
(symbols) for T=0.3 obtained at various pulling velocities for the good folder together with 
the corresponding WHAM estimate $f_W(\zeta)$ (dashed lines).
As a first point, we notice that the estimated FEL collapses towards
$f_W(\zeta)$ as the pulling velocity decreases.
In particular, for the good folder the asymptotic shape is 
reached for small $\zeta$-values at a somehow larger velocity (namely,
for $\zeta < 10$ already for $v_p = 5 \times 10^{-4}$) than at 
larger $\zeta$. In particular, to
reproduce $f_W(\zeta)$  up to $\zeta_{trans}$ the pulling should be
performed at $v_p = 5 \times 10^{-6}$.
Moreover, referring to Fig.~\ref{gf_eje_t0.3}, it is possible to identify the structural transitions (STs) induced
by the pulling experiment. As shown in Fig.~\ref{gf_eje_t0.3} (b), the asymptotic $f_J(\zeta)$ profile exhibits a 
clear minimum in correspondence of the end-to-end distance of the 
NC (namely, $\zeta_0 \sim 1.9$). In more detail, up to $\zeta \sim 5.6$, the protein remains in 
native-like configurations characterized by a $\beta$-barrel made up of 4 strands, while the 
escape from the native valley is signaled by the small dip at $\zeta \sim 5.6$ and it is
indicated as ST1 in Fig. \ref{gf_eje_t0.3} (b). 
This ST has been firstly identified in \cite{lacks} by analyzing the 
the potential energy of ISs measured during a mechanical unfolding 
(numerical) experiment. In particular, Lacks \cite{lacks} identifies
this transition as an irreversible transition, in the sense that
above this transition it is no more sufficient to reverse the stretching
to recover the previously visited configurations \footnote{Please notice
that we observe this transition at $\zeta \sim 5.6$ and not at
$\zeta = 4.782$ as Lacks has reported, since we are considering the free energy
profile at $T=0.3$, while Lacks' analysis concerns potential energies of the ISs.
Our inspection of the average potential energies estimated during the pulling
experiments and reported in Fig. \ref{entropy_energy}(a) confirms this small mismatch.}.

\begin{figure}[t]
\includegraphics[draft=false,clip=true,height=0.34\textwidth]{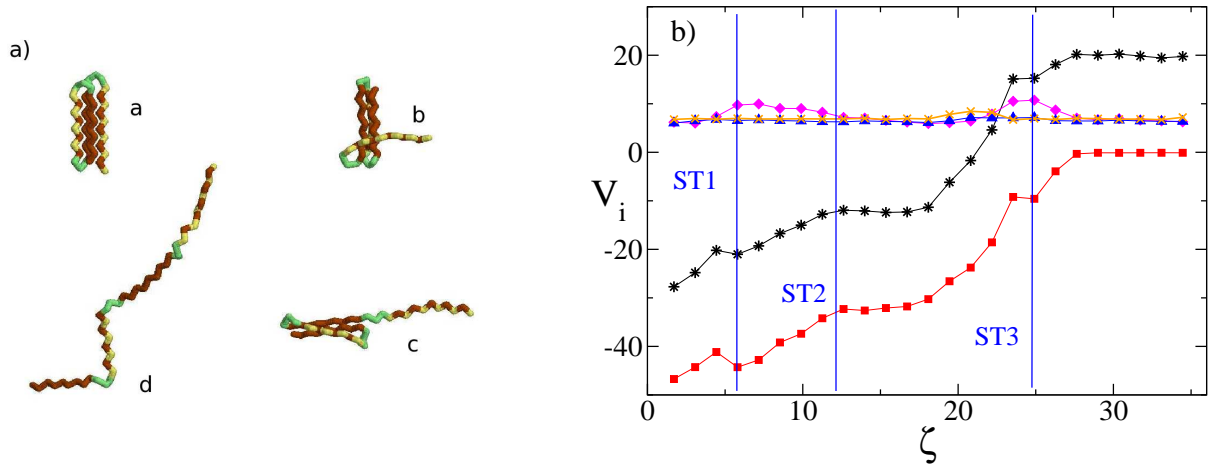}
\includegraphics[draft=false,clip=true,height=0.34\textwidth]{f6b}
\caption{(Color online) (a) Pulled configurations of the good folder at $T=0.3$:
the NC (a) has $\zeta_0 \sim 1.9$;
the others are characterized by $\zeta=6.8$ (b), $\zeta=16.8$ (c), and $\zeta=27.1$ (d).
The beads of type $N$, $B$, and $P$ are colored in green,
red and yellow, respectively. (b)  Potential energies contributions 
as a function of the end-to-end distance $\zeta$ estimated during a pulling
experiment with speed  $v_p = 5 \times 10^{-6}$ and obtained by
averaging over 28 different realizations at  $T=0.3$. (Black) Stars
indicate the entire potential energy $V$, (orange) crosses $V_1$, (blue) triangles
$V_2$, (magenta) diamonds $V_3$, and (red) squares $V_4$. 
The (blue) vertical solid lines indicate the transitions previously discussed in the text.
}
\label{config.gf}
\end{figure}

For $\zeta > 6 $ the configurations are
characterized by an almost intact core (made of 3 strands) plus a stretched 
tail corresponding to the pulled fourth strand (see (b) in Fig. \ref{config.gf}(a)).
The second ST amounts to pull the strand $(PB)_5P$ out of the barrel.
In the range $13 < \zeta < 18.5$ the curve $f_J(\zeta)$ appears as essentially flat,
thus indicating that almost no work is needed to completely stretch the tail once detached from 
the barrel (see configuration (c) in Fig. \ref{config.gf}(a)). The pulling of the third strand (that is part 
of the core of the NC) leads to a definitive destabilization of the $\beta$-barrel. This transition is denoted 
as ST3 in Fig. \ref{gf_eje_t0.3} (b). The second plateau in $f_J(\zeta)$ corresponds to protein structures made 
up of a single strand (similar to (d) in Fig. \ref{config.gf}(a)). 

To distinguish between entropic and energetic costs associated to each ST we have also evaluated
separately the potential energy contributions $V_i$ ($i=1,\dots,4$) during the pulling experiment, 
these data are reported in Fig. \ref{config.gf}(b). From the figure it is clear that the variation of the 
potential energy during the stretching is essentially due to the Lennard-Jones term $V_4$, while
the other terms contribute to a much smaller extent, at least up to $\zeta \sim 35$.
The transition ST1 has essentially only energetic costs, since $\Delta f = 7(1)$ and
the potential energy varies almost of the same amount, in particular $\Delta V \sim \Delta V_4 = 8(1)$.
The other transitions instead have not negligible entropic costs, since the free energy barrier heights
associate to ST2 and ST3 are 10(1) and 29(2), respectively; while the corresponding potential energy barriers
are higher, namely $\Delta V = 16(1)$ for ST2 and $\Delta V = 43(1)$ for ST3. The complete
stretching of the protein up to $\zeta=35$ has a free (resp. potential) energy cost corresponding to
$\Delta f = 30(2)$ (resp. $\Delta V = 49(1)$).
Above $\zeta \sim 35$, while the Lennard-Jones and dihedral contributions vanish,
the final (almost quadratic) rise of the free energy is due to the harmonic and angular
contributions, since we are now stretching bond distances and angles beyond their equilibrium values.
Due to computational constraints and to the fact that this part of the FEL is not particularly relevant,
the reconstructions at the lowest velocities and the WHAM estimations have been not performed
for these large $\zeta$-values.

\begin{figure}[t]
\includegraphics[draft=false,clip=true,height=0.34\textwidth]{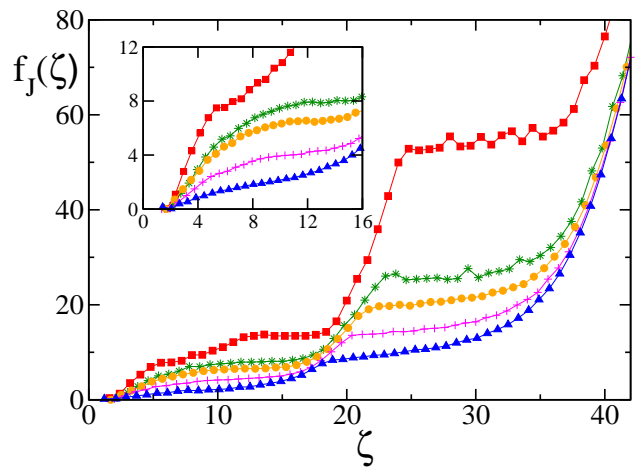}
\caption{(Color online) Free energy profiles $f_J(\zeta)$ obtained with the EJE for good folder 
at various temperatures:
$T=0.2$ (red squares), $0.4$ (green  stars), $0.5$ (orange circles), 
$0.6$ (magenta plus) and $0.7$ (blue triangles). 
In the inset, an enlargement is reported at small 
$\zeta$. Data refer to $v_p=5 \times 10^{-4}$.
The number of different realizations performed to estimate the averages
at the different temperatures ranges between $160$ and $250$.}
\label{gf_eje_varieT}
\end{figure}

In Fig. \ref{gf_eje_varieT} the reconstruction of the FEL obtained at various temperatures is
shown. For temperatures around $T_f$ one still observes a FEL resembling the one found for $T=0.3$, while
by increasing the temperature the dip around $\zeta \sim 6 - 7$ (associated to ST1) disappears and the 
heights of the other two barriers reduce.
By approaching $T_\theta$ the first plateau, characterizing the transition from the NC to configurations
of type (c), essentially disappears, and it is substituted by a monotonous increase of $f_J(\zeta)$.
This suggests that 4 stranded $\beta$-barrel configurations coexist with partially unfolded ones. 
Above $T_\theta$  only one barrier remains indicating that at these temperatures the protein
unfolds completely in one step process.

\begin{figure}[t]
\includegraphics[draft=false,clip=true,height=0.34\textwidth]{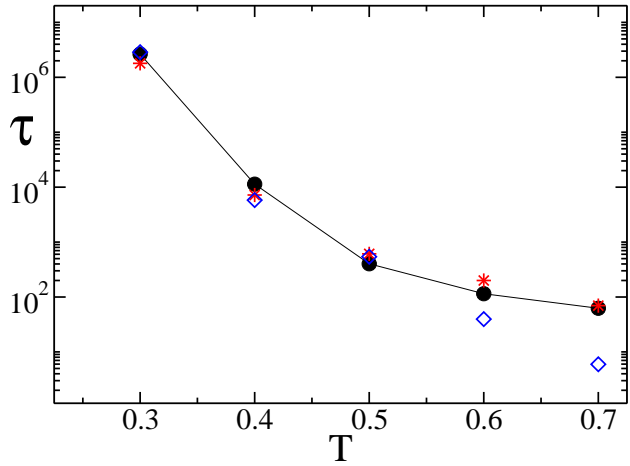}
\caption{(Color online) Average unfolding times $\tau$ for the GF at various
temperatures corresponding to $\zeta_{th}=4$. Filled (black) circles denote the numerical data, 
the estimations obtained via eq. (\ref{TST}) and eq. (\ref{smo}) are
represented by empty (blue) diamonds and (red) stars, respectively.
The arbitrary scaling factor entering in eq. (\ref{smo}) (see text) 
has be set equal to 8.  The average times have been estimated over
$100,000$ -- $200,000$ unfolding events for $T=0.7$ and 0.6,
$12,000$ events at $T=0.5$ and as few as $200$ and $60$ events
at the lowest temperatures, namely $T=0.4$ and $0.3$.
}
\label{fpt}
\end{figure}

The connection between dynamical properties of the system and the free energy profile is
still an open problem. In particular, the relationship between the unfolding times and the
free energy barriers has been previously discussed in Ref. \cite{west} for proteins and more recently
the same problem has been addresses for Ising-like lattice protein model in Ref. \cite{zamparo}.
We have estimated average first passage times $\tau$ via USs by recording the time
needed to the protein to reach a certain end-to-end threshold $\zeta_{th}$ once it
starts from the NC at different temperatures. Our data, reported in Fig. \ref{fpt},
clearly indicate that at low temperatures the simple result of the transition
state theory~\cite{langer,review,zwanzig}, namely
\begin{equation}
\tau = \frac{{\rm e}^{\Delta f/T}}{T} \qquad ,
\label{TST}
\end{equation}
where $\Delta f = f(\zeta_{th}) - f(\zeta_0)$, is in very good agreement with the numerics.
However, at high temperatures the agreement worsens. Therefore, in order to
take in account all the details of the free energy profile and not only the barrier height,
we have generalized a result of the Smoluchowski theory for the
escape of a particle from a potential well~\cite{zwanzig} as follows~\cite{zamparo}
\begin{equation}
\tau \propto \frac{1}{T} \int_{\zeta_0}^{\zeta_{th}}
\enskip dy \enskip {\rm e}^{f(y)/T} \int_{\zeta_0}^y dz {\rm e}^{-f(z)/T}
\label{smo}
\end{equation}
where the potential energy has been substituted by the free energy profile. The estimation
obtained via eq. (\ref{smo}) compare well with the numerical results at all the
considered temperatures, unfortunately apart an arbitrary scaling factor common to all 
the temperatures that we are unable to estimate (see Fig. \ref{fpt}).

\subsection{Bad Folder}

\begin{figure}[t]
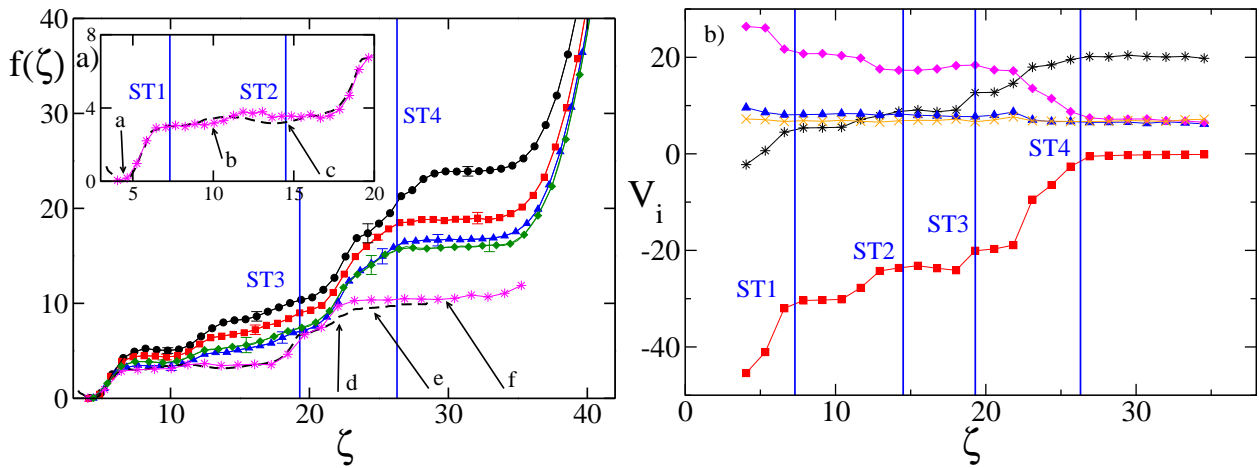

\includegraphics[draft=false,clip=true,height=0.34\textwidth]{f9a}
\includegraphics[draft=false,clip=true,height=0.34\textwidth]{f9b}
\caption{(Color online) (a) Free energy profiles $f_J$ for the bad folder as a function 
of the end-to-end distance $\zeta$, 
obtained with the EJE for various pulling velocities: from top to bottom
$v_p=5 \times 10^{-4}$ and $160$ realizations (black circles), 
$2 \times 10^{-4}$ and $200$ realizations (red squares),  
$1 \times 10^{-4}$ and $200$ realizations (blue triangles),
$5 \times 10^{-5}$ and $100$ realizations (green diamonds), 
$5 \times 10^{-6}$ and $28$ realizations (magenta stars). 
The WHAM estimate $f_W(\zeta)$ is also shown (black dashed line).
In the inset an enlargement of the curve
at low $\zeta$ for $v_p=5 \times 10^{-6}$ is reported
together with $f_W(\zeta)$.
Data have been obtained at $T=0.3$.
(b)  Potential energies contributions
as a function of the end-to-end distance $\zeta$ estimated during a pulling
experiment with velocity $v_p = 5 \times 10^{-6}$ and obtained by
averaging over 28 different realizations at  $T=0.3$. Black stars
indicate the entire potential energy $V$, (orange) crosses $V_1$, (blue) triangles
$V_2$, (magenta) diamonds $V_3$, and (red) squares $V_4$. The (blue) solid
lines indicate the transitions discussed in the text.}
\label{Bf_eje_t0.2}
\end{figure}

In Fig. \ref{Bf_eje_t0.2}(a) are reported the free energy profiles
$f_J(\zeta)$ reconstructed via the EJE at $T=0.3$ for different pulling speeds
(symbols) together with the estimated $f_W(\zeta)$ (dashed line), 
as in the case of the GF one observes a collapse to the equilibrium 
FEL (represented by $f_W(\zeta)$) for a sufficiently small speed. 
In particular, at $v_p= 5 \times 10^{-6}$ a reasonably good 
agreement between $f_J$ and $f_W$ is already achieved.

For the BF the mechanically induced unfolding transition 
are less clearly identifiable from the inspection of the free energy 
profile for two reasons. Firstly, for the BF not only the LJ interactions play a role in 
the STs but also the dihedral terms: these two terms contribute with opposite signs 
to the whole potential energy thus partially canceling each other. Moreover, 
as we will show in the following the main contribution to the free energy is
due to entropic terms.  Therefore,
in order to identify the STs it is better to consider the distinct average profile 
of the single potential contributions $V_i$ ($i=1,\dots,4$) reported in Fig. \ref{Bf_eje_t0.2}(b).
In particular, the most relevant is the Lennard-Jones term $V_4$, due to the
stabilizing effect of the hydrophobic interactions on the protein structure.
From the inspection of $V_4$, at least four different STs can be single out,
occurring at $\zeta \sim7.3$, 14.5, 19.3, and 26.3, respectively.

\begin{figure}[t]
\includegraphics[draft=false,clip=true,height=0.34\textwidth,angle=0]{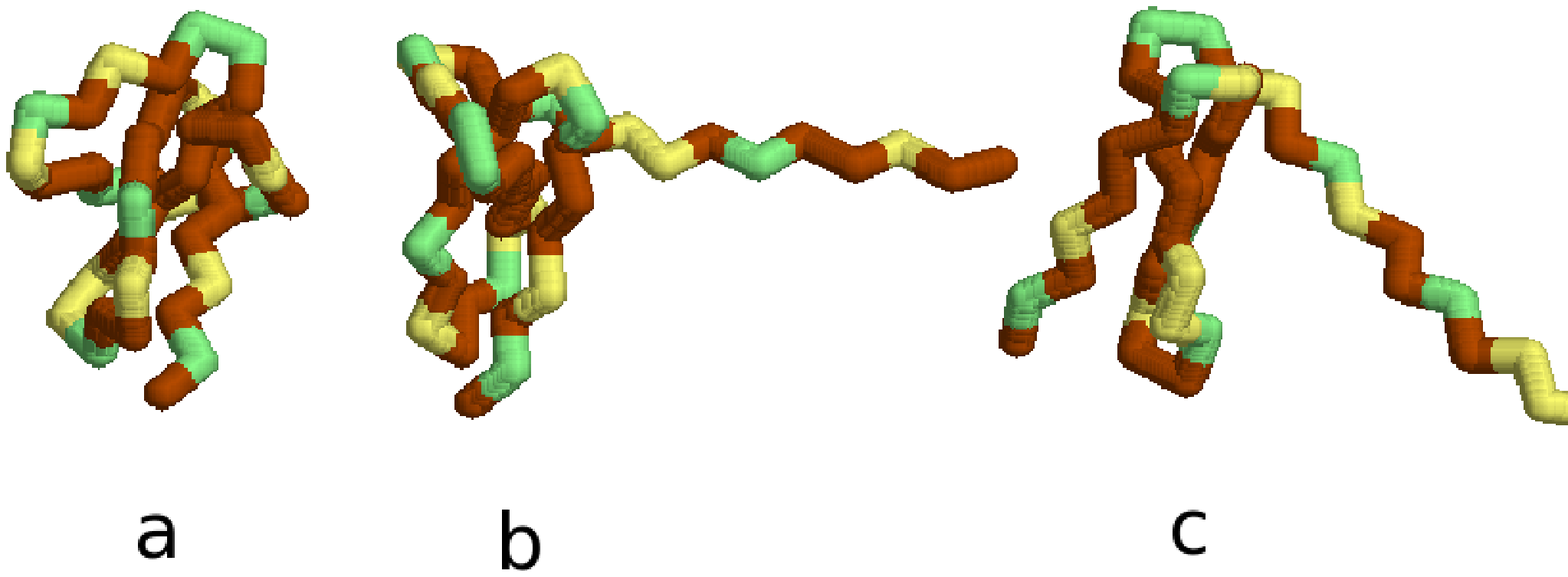}
\includegraphics[draft=false,clip=true,height=0.34\textwidth,angle=0]{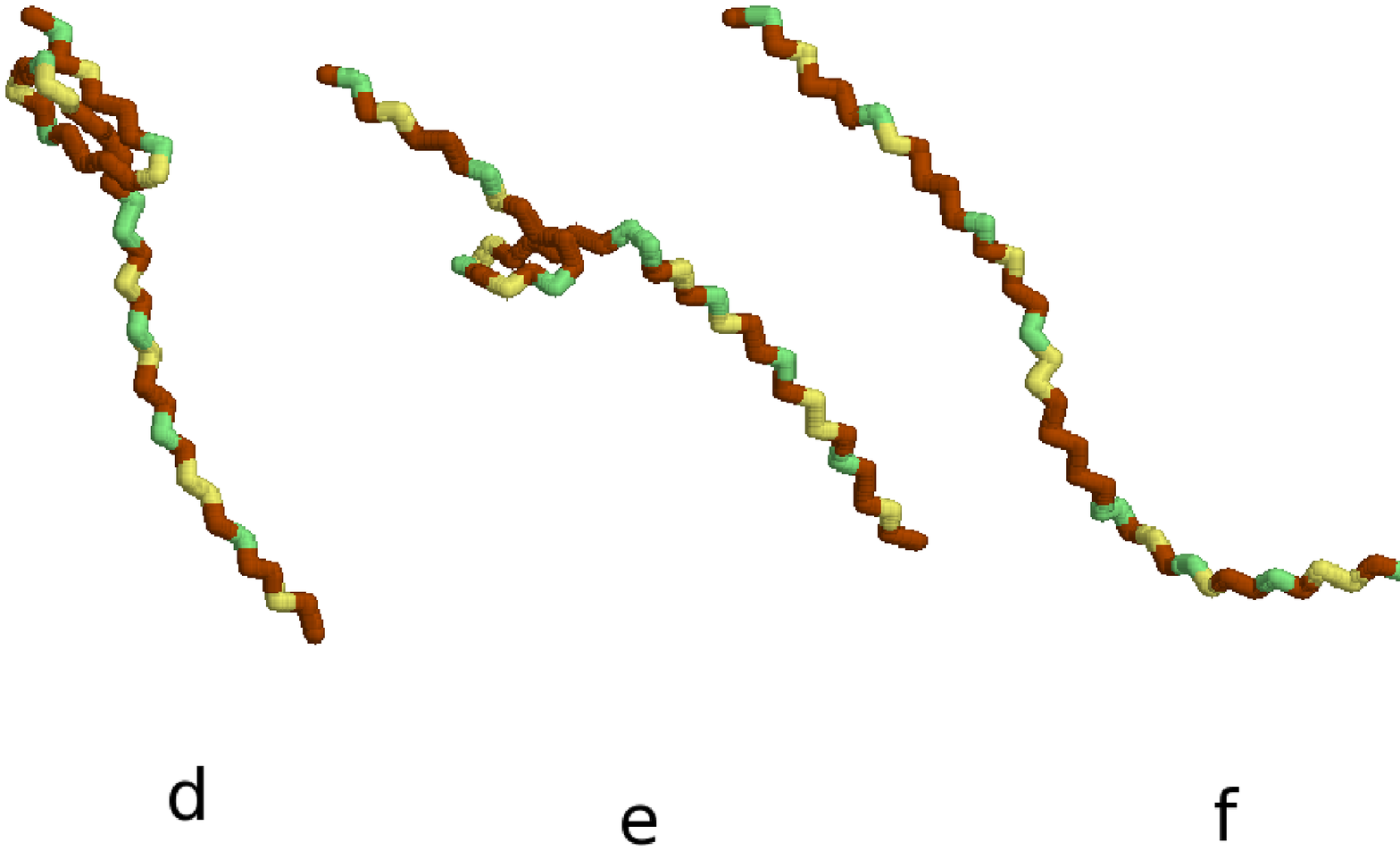}
\caption{(Color online) Pulled configurations of the bad folder at $T=0.3$:
the reported configurations refer to $\zeta_0=4.7$ (NC) (a),
$\zeta=9.9$ (b), 14.5 (c), 22.1 (d), 24.6 (e), and 29.7 (f).
}
\label{config.bf}
\end{figure}

The first transition amounts to pull the last part of the tail out of
the NC, namely the 6th and 5th strand that we have previously identified.
To this ST is associated a free energy increase of 3.1(5) and
a potential energy variation of 8.0(5), once the ST1
is completed the protein assumes the configuration (b)
shown in Fig. \ref{config.bf}. ST2 consists in pulling out from 
the compact configuration the whole tail (therefore to detach also 
the 4th strand) and leaving the protein in a configuration composed
by the core (represented by the first three strands) plus a long tail
(see configuration (c) in  Fig. \ref{config.bf}). The entropic contributions
to ST2 is quite relevant since to pass from the NC to (c)
the free energy increases of 3.8(5), 
while the associated potential energy variation is almost the triple, i.e. 11.5(5). The third transition amounts
to detach the first $\beta$-strand ($BNBPB_{3}NP$) from the core
and this operation has much greater costs with respect to the previous STs,
namely, $\Delta f = 7.0(5)$ and $\Delta V= 15(1)$. The complete opening of
the core structure (now made only of the second and third strand)
occurs at $\zeta \sim 27$ amounting to a total free (resp. potential) energy barrier
to overcome of height $11(1)$ (resp. $23(1)$). 
At variance with the GF case, for the BF the entropic costs 
are never negligible and instead they always amount at least at 
the half of the potential energy contributions in all the four examined transitions.
Finally, analogously to the GF for $\zeta > 35$ the LJ and dihedral contributions
essentially vanish and the free energy increase is due to the harmonic and
angular terms, only.

\begin{figure}[t]
\includegraphics[draft=false,clip=true,height=0.34\textwidth]{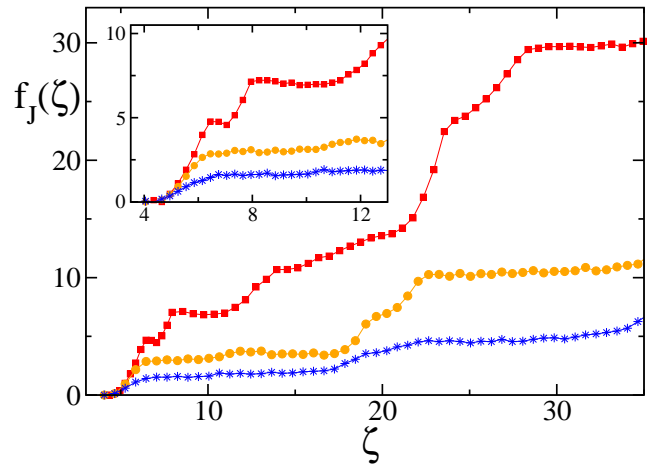}
\caption{(Color online) Free energy profiles $f_J(\zeta)$ obtained via the EJE for bad 
folder at three temperatures: namely,  $T =0.2 $  (red squares), $T=0.3$ (orange circles) ,
$T=0.4$ (blue stars). In the inset an enlargement is reported at small $\zeta$. 
Data refer to pulling velocity $v_p = 5 \times 10^{-6}$ 
and the averages are performed over $28$ samples of the same protocol.}
\label{bf_varieT}
\end{figure}

In Fig. \ref{bf_varieT} the reconstruction $f_J$ of the FEL for the bad folder 
is reported at three temperatures below $T_\theta$. As one can notice the bad folder 
exhibits at comparable temperatures
much lower free energy barriers, indicating that the NC and the 
partially folded structures are less stable, 
with respect to the GF. This is reflected also
in the value of $T_\theta$ that has a smaller value with respect to the
GF: namely, 0.46 for BF and 0.65 for GF. By increasing $T$ 
the heights of the free energy barriers rapidly decrease and the various
STs become less clearly defined. Moreover, the FEL of the BF at the lower examined
temperature ($T=0.2$) reveals, besides the absolute minimum (corresponding to the NC), 
other two local minima at $\zeta \sim 7$ and $\zeta \sim 11$. This indicates 
that, at variance with the GF, the BF can remain trapped even at $T \sim T_f$, 
for some finite time, in intermediate (misfolded) states far from the NC.

\section{Inherent structure landscape}
\label{seven}

\begin{figure}[t]
\includegraphics[draft=false,clip=true,height=0.4\textwidth]{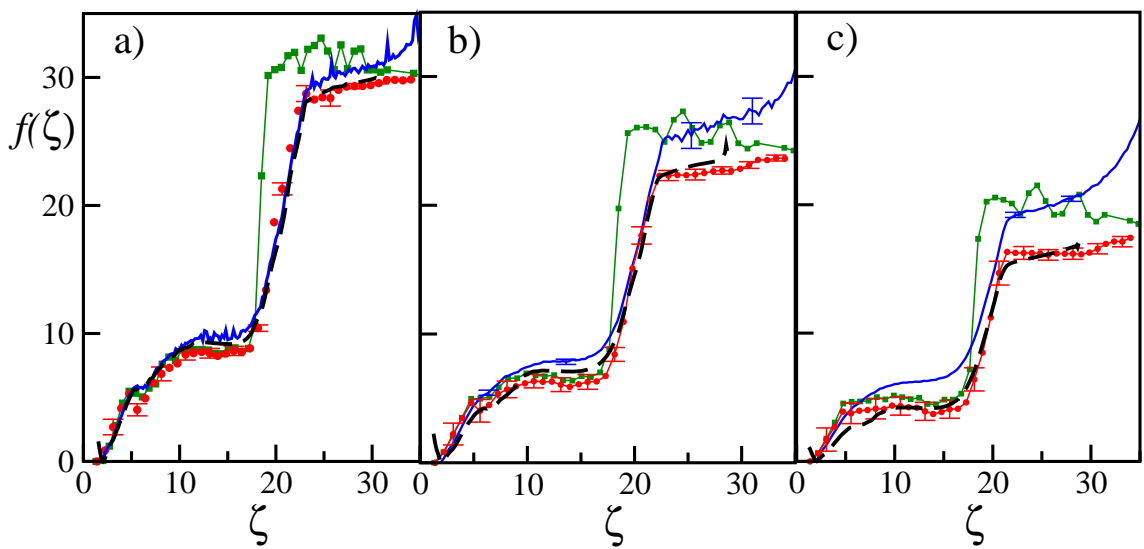}
\caption{(Color online) Free energy profiles $f_J$ 
(blue solid lines) as a function of $\zeta$ for various temperatures 
for the good folder: a) $T=0.3$ for $v_p = 5\times 10^{-6}$ and $28$ repetitions; b) $T=0.4$ for
$v_p = 5\times 10^{-4}$ and $240$ experiments; c) $T=0.5$ 
for $v_p = 5\times 10^{-4}$ and $240$ repetitions.
The (black) dashed lines refer to the WHAM estimation $f_W(\zeta)$, (green) squares to $f_{IS}(\zeta)$ 
obtained by employing the TDB and  (red) circles to $f_{IS}(\zeta)$ obtained by employing
the ISs in the PDB for each considered $T$.}
\label{gf_varT_cfr_pul_mi_tot}
\end{figure}

In this section we compare the reconstructions of the FEL for the good and bad folder obtained 
via the EJE and the IS approach with the WHAM equilibrium estimation. As already explained in Sect. III,
we have created two IS data banks : the thermal data bank (TDB) obtained by 
performing equilibrium canonical simulations and the 
pulling data bank (PDB) by mechanically unfolding the protein. 
In Fig.~\ref{gf_varT_cfr_pul_mi_tot} is reported for the GF the comparison, at three temperatures,
between the estimate $f_W(\zeta)$ with $f_{IS}(\zeta)$ and the $f_J(\zeta)$, obtained via the EJE 
reconstruction. 
The results reveal an astonishingly good coincidence between $f_W(\zeta)$ and $f_{IS}(\zeta)$,
obtained by employing the PDB, at all the examined temperatures. For what concerns the
EJE reconstructions: at $T=0.3$ $f_J(\zeta)$ is essentially in good agreement with the other
two estimations, while at higher temperatures the $f_J$ curves slightly overestimate
the equilibrium free energy $f_W$ for $\zeta > 10$.  This discrepancy is probably due to a non 
complete convergence of the EJE approach at the considered pulling velocities, smaller velocities
are required to recover the equilibrium profile at all th end-to-end distances.

The further comparison reported in Fig.~\ref{gf_varT_cfr_pul_mi_tot} between the IS
reconstructions obtained via the TDB and the PDB indicates a perfect coincidence up to $\zeta \sim 17$.
On the contrary, during the last stage of the unfolding process the two $f_{IS}$ differ: the TDB FEL is
steeper than the PDB one. This suggests that during the mechanical unfolding the protein can
easier reach states with low energies, even at large $\zeta$. These states have a very low probability
to be visited during thermal equilibrium dynamics. However, at $T=0.3$
the value of the barrier to overcome and that of the final plateau are quite similar to those of the PDB FEL,
while at higher temperatures the final energy plateaus of the TDB FEL
are slightly larger than the $f_W$-plateaus. The reason of these discrepancies is
related to the fact that, despite the high number of IS forming the TDB, this data bank 
is far from containing all the relevant ISs, in particular those associated to high $\zeta$-values
are lacking. It should be remarked that the IS conformation with the maximal end-to-end distance is the 
all {\it trans}-configuration, corresponding to $\zeta_{trans}=35.70$,
therefore the IS approach does not allow to evaluate the FEL for $\zeta > \zeta_{trans}$.
For the GF, we can safely affirm that the out-of-equilibrium process 
consisting in stretching the protein is more efficient to investigate the FEL, 
since a much smaller number of ISs are needed to reliably reconstruct it, 
as reported in Table \ref{IS_PDB}.

\begin{figure}[t]
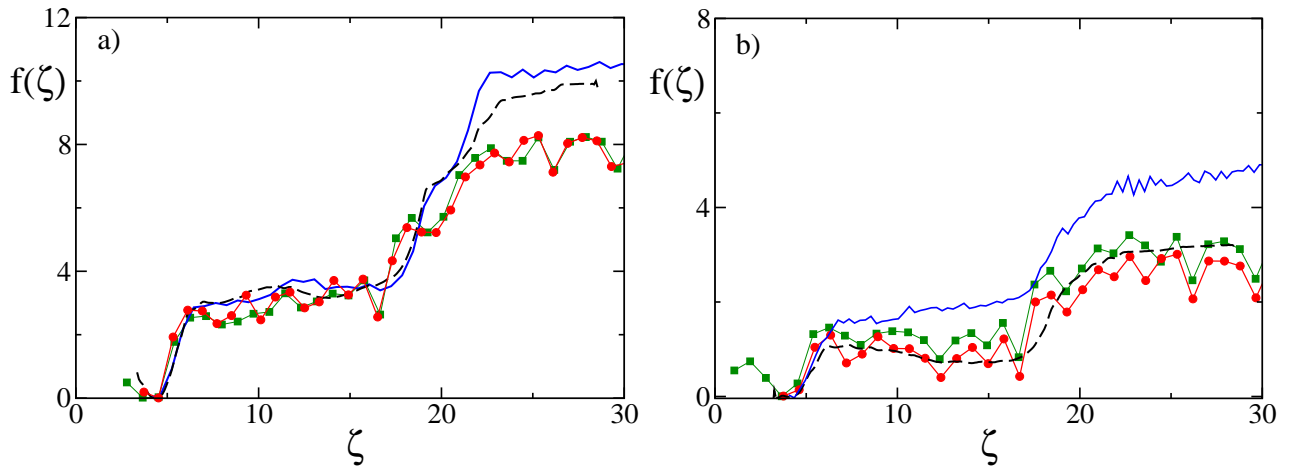

\includegraphics[draft=false,clip=true,height=0.34\textwidth]{f13a}
\includegraphics[draft=false,clip=true,height=0.34\textwidth]{f13b}
\caption{(Color online) Free energy profiles $f_J$ as a function of $\zeta$ 
for various temperatures for the bad folder:
a) $T=0.3$ and b) $T=0.4$. The data refer to a pulling velocity
$v_P=5 \times 10^{-6}$ and $28$ repetitions of the same pulling protocol.
The symbols are the same as in Fig. \ref{gf_varT_cfr_pul_mi_tot}. 
}
\label{Bf_varT_cfr_pul_mi_tot}
\end{figure}

The comparison for the BF case is reported in Fig. \ref{Bf_varT_cfr_pul_mi_tot}
at $T=0.3$ and 0.4. Also in this case the $f_W(\zeta)$ and $f_{IS}(\zeta)$ 
essentially coincide, apart at $T=0.3$ and $\zeta > 20$ where $f_W$ is slighty higher than
$f_{IS}$.  In this case the agreement between the two IS reconstructions is quite good at both
the considered temperatures and for all $\zeta$-values.  As far as the EJE reconstructions
are concerned, at the employed pulling velocity (namely, $v_P=5 \times 10^{-6}$)
$f_J$ can be considered as asymptotic at $T=0.3$, while probably at $T=0.4$ is still
slightly overestimating $f_W$, but please notice the really small range of the
free energy scale reported in Fig. \ref{Bf_varT_cfr_pul_mi_tot}(b) with respect to the GF.

\begin{figure}[t]
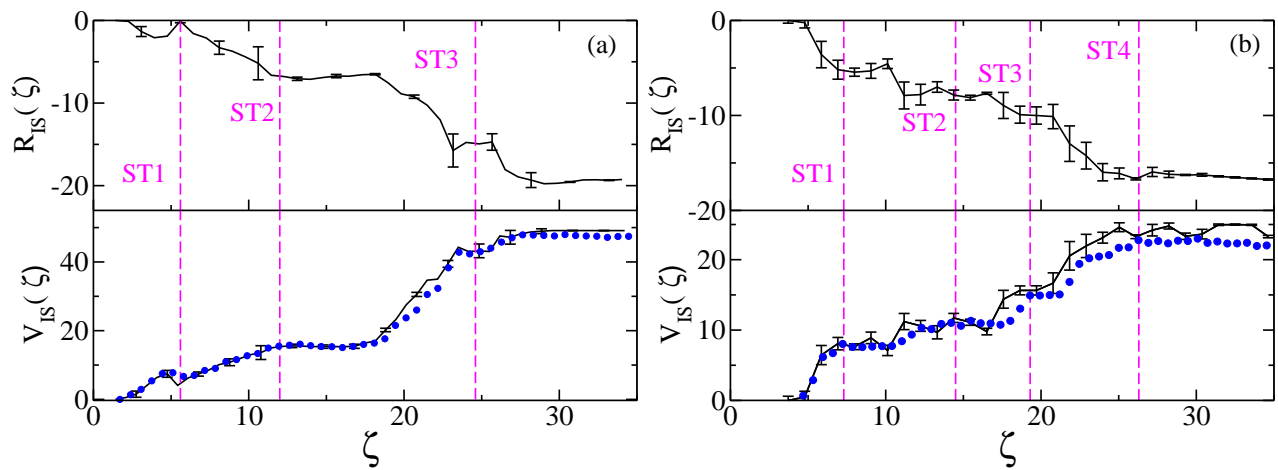

\includegraphics[draft=false,clip=true,height=0.34\textwidth]{f14a.eps}
\includegraphics[draft=false,clip=true,height=0.34\textwidth]{f14b.eps}
\caption{(Color online) Reconstructed $V_{IS}(\zeta)$ (lower panel) and $R_{IS}(\zeta)$ (upper panel) 
for good folder (a) and bad folder (b) by employing ISs in the PDB at $T=0.3$. In the lower panel the blue dotted line
refers to the average potential energy evaluated during the corresponding pulling experiments (this has been 
already reported in Fig. \ref{config.gf}(b) for the GF and in Fig. \ref{Bf_eje_t0.2}(b)) for the BF. 
Please notice that the data have been vertically translated 
in order to have zero energy at the NC.}
\label{entropy_energy}
\end{figure}

Furthermore, from the IS analysis by employing Eq. (\ref{VIS_RIS}) we can obtain an estimate of the 
profiles of the potential and vibrational free energies $V_{IS}(\zeta)$ and $R_{IS}(\zeta)$, respectively. 
From the latter quantity, the entropic costs associated to the various unfolding stages can be estimated.
As shown in Fig.~\ref{entropy_energy}(a), for the GF at $T=0.3$, the structural transitions ST2 and ST3
previously described correspond to clear "entropic" barriers, while the ST1 transition 
has only energetic costs since $\Delta R_{IS} \sim 0$. This last result is in good agreement with 
the previously reported EJE analysis. For what concerns the other two transitions, ST2 (resp. ST3)
is associated to a decrease $\sim 6(1)$ (resp. $15(2)$) of $R_{IS}(\zeta)$ once more in agreement with
the EJE reconstruction. The complete opening of the protein is associated to a barrier
$\Delta R_{IS}(\zeta) = 20(2)$, while the analysis reported in Sect.\ref{six}A indicates an entropic
barrier to overcome corresponding to $\sim 19(2)$. These results suggest that for the good folder
the entropic contributions to the free energy are essentially of the vibrational type.
Moreover, the reconstructed potential energies $V_{IS}(\zeta)$ are in very good agreement with the average 
potential energy evaluated during the corresponding pulling experiments as shown in Fig.~\ref{entropy_energy}(a).

Finally, one can try to put in correspondence the three unfolding stages previously 
discussed for the GF with thermodynamical aspects of the protein folding. In particular,
by considering the energy profile $V_{IS}(\zeta)$, an energy barrier $\Delta V_{IS}$  
and a typical transition temperature $T_t = (2 \Delta V_{IS})/(3N)$
can be associated to each  of the STs. The first transition ST1 corresponds to a barrier 
to overcome $\Delta V_{IS} = 8(1)$ and therefore to $T_t = 0.11(1)$, that, within error bars, 
coincide with $T_g$. For the ST2 transition the barrier to overcome is $\Delta V_{IS} = 16(1)$ and this is 
associated to a temperature $T_t \simeq 0.23(2)$ (slightly smaller than $T_f$). 
At the ST3 transition $\Delta V_{IS} = 43(2)$ corresponding to $T_t = 0.62(2)$, while
the energetic cost to completely stretch the protein is $50(2)$ with an associated transition 
temperature $T_t = 0.72(2)$: the $\theta$-temperature ($T_\Theta = 0.65(1)$) is well bracketed
within these two transition temperatures. At least for the GF , our results indicate that the observed STs
induced by pulling can be put in direct relationship with the thermal transitions usually identified 
for the folding/unfolding process.

Also for the BF the IS approach is able to well reproduce not only the average potential energy during the
pulling experiment, as clearly shown in Fig.~\ref{entropy_energy}(b), but also to provide
a good estimate of the ``entropic'' barriers associated to the
structural transitions. In particular, at $T=0.3$ the vibrational free energy barriers to 
overcome are $\Delta R_{IS} = 5.3(5)$ at ST1, 8(1) at ST2, 10(1) at ST3 and 16(1) at ST4.
These values are in reasonably good agreement with those previously obtained from the EJE reconstruction,
apart at ST3 and ST4, where the analysis performed in Sect.\ref{six}B indicates entropic
barriers to overcome corresponding to $\sim 8(1)$ and $\sim 12(2)$, respectively.
These underestimations at large $\zeta$-values are probably due to the fact that 
at this temperature the estimated $f_J$ has not reached its asymptotic shape
at the employed velocity.

As already previously pointed out, the entropic contributions for the BF are more relevant than for the GF:
e.g while the ST2 transition is clearly visible by the potential energy inspection
it is almost absent by looking to the free energy profile (compare the data reported 
Fig. \ref{Bf_eje_t0.2}(a) and (b)). Therefore we cannot expect to infer information on the thermal 
transitions from the knowledge of the potential energy barriers at the STs, as done for
the GF. Indeed the estimated transition temperatures $T_t$ for the four examined structural
transitions give values not corresponding to any of the relevant temperatures reported in
Table \ref{transition_temperatures} for the BF.

\begin{figure}[t]
\includegraphics[draft=false,clip=true,height=0.4\textwidth]{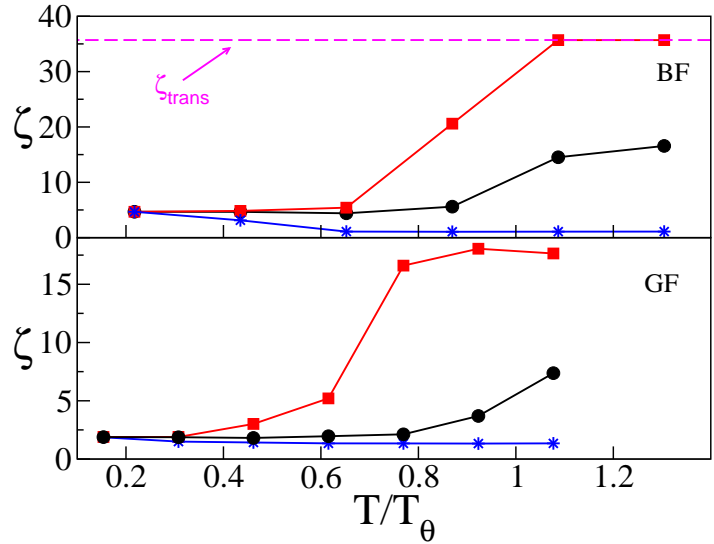}
\caption{(Color online) End-to-end distance of the ISs estimated
during USs at various temperatures: (black) circles represent the
average value; (blue) stars the minimal value; and (red) squares the maximal one.
The upper panel refer to the BF and the lower one to the GF. The horizontal magenta
dashed line indicates the $\zeta_{trans}$-value. For the GF (resp. BF) trajectories
of duration $t \sim 100,000 - 500,000$ (resp. $t \sim 50,000 - 250,000$) have
been examined to obtain the ISs at constant time intervals $\Delta t = 5$.
}
\label{ete_GF_BF}
\end{figure}

To better understand this difference we have performed USs for the GF and BF for
$T_g \lesssim T \lesssim T_\theta$ and we have estimated the average, the minimal and the maximal $\zeta$
associated to the visited ISs. The corresponding data are reported in Fig.~\ref{ete_GF_BF}.
While for the GF the minimal value remains essentially $\zeta_0$ for all the temperatures and
the maximum $\zeta$ increases smoothly up to $\sim 18$ at $T=T_\theta$, the
dependence of the minimal and maximal $\zeta$-values on $T$ are more dramatic for the BF.
Up to the temperatures $T \sim 0.5 \times T_\theta$, average , minimal and maximal $\zeta$-values
almost coincide
indicating that the protein is still confined around the NC, please remember that for the
BF $T_g = 0.58 \times T_\theta$. As soon as $T > 0.6 \times T_\theta$
the maximum grows abruptly and reach the upper bound corresponding to $\zeta_{trans}$ already at
$T \sim T_\theta$, on
the other hand the minimum value decreases indicating that at higher temperatures the protein
can access basins of ISs with end-to-end distance lower than $\zeta_0$. This last result
indicates that there is not a clear monotonic correspondence between the temperature increase
and the achievable protein extensions. Moreover, the fact that the protein can easily
attain also extremely stretched configurations at not too high temperatures suggests
that in the case of the BF the protein can easily escape form the native valley and reach
any part of the phase space, while for the GF the accessible IS configurations are
much more limited at comparable temperatures. All this amounts to say that the end-to-end
distance cannot be considered as a good reaction coordinate for the BF.

\section{Concluding remarks}
\label{eight}

In conclusion, we can safely affirm that the reconstructions of the
free energy landscape as a function of the end-to-end distance 
in terms of the ISs,  obtained via out-of-equilibrium 
mechanical unfolding of the heteropolymers, are in very good agreement 
with the equilibrium weighted histogram estimate for the good and bad folder
sequences at all the examined temperatures. In particular, this result
indicates that the harmonic approximation employed to estimate the vibrational 
term (\ref{fvib}) is quite good for temperatures in the range
$[T_f;T_\theta]$, as already pointed out in \cite{kim-keyes}
by considering the average potential energy.
Moreover, the EJE reconstructions
of the free energy profile compare quite well with the other two approaches
for sufficiently low pulling velocities.
For the good folder, the quality of the free energy landscape 
reconstruction via the extended Jarzinsky equality can be well appreciated by 
stressing that from pure structural information about the landscape 
a good estimate of dynamical quantities, like the unfolding times from the native 
configuration, can be obtained. 

Furthermore, for the good folder the information obtained by the equilibrium FEL
both with the EJE and the IS methodologies can be usefully combined to
give substantiated hints about the thermal unfolding.
In particular the investigation of the ISs allows us to give an estimate
of the (free) energetic and entropic barriers separating the
native state from the completely stretched configuration.
These barriers are associated to the structural transition induced
by the protein manipulation and for the good folder they can put in 
direct relationship with the thermal transitions usually identified during 
folding/unfolding processes.

On the other hand for the bad folder the end-to-end distance 
appears not to represent a good reaction coordinate, since mechanical
and thermal unfolding seem to follow different paths.
In other terms the unfolding process
for the good folder consists of many small successive 
rearrangements of the NC, which are well captured by the
distribution of the corresponding ISs on the landscape. 
While for the bad folder the thermal unfolding can involve also large 
conformational rearrangements, thus implying jumps from one valley to another of the
landscape associated to large variations in the end-to-end distance, that
cannot be well reproduces by the mechanical stretching of the heteropolymer.
Future work on more realistic heteropolymer models is needed to 
clarify if the observed features, distinguishing good folders from bad folders,
can be really considered as a specific trademark of proteins.

A drawback of the EJE reconstruction is that extremely small 
velocities or an extremely large number of repetitions of the protocol 
are needed to achieve the collapse towards the equilibrium profile, thus 
rendering the implementation of the method quite time consuming.
However, new optimized methods to obtain the asymptotic FEL, 
by combining the Jarzinsky equality with the Crooks' path ensemble average theorem, 
have been recently published \cite{chelli,bidirectional}
and it will be definitely worth to test their performances 
in the next future with respect to complex landscapes, like those of heteropolymers \cite{future}.

As a final point, we would like to remember that, in the context of glassy systems, 
the concept of ISs has been critically compared to that of pure states \cite{biroli}, 
the latter being local minima of the free energy landscape, while the ISs are minima of the potential 
energy, as discussed above.
The relevance of the pure states for protein folding has been recently stressed in Ref. \cite{svezia},
where it has been shown for a fibronectin domain that pure states can be put in direct 
correspondence with unfolding intermediates observable during mechanical pulling.
However, in the present paper we have been only interested in how the FEL, 
which is the only thermodynamical relevant function, together with the corresponding pure
states, can be obtained by employing a suitably chosen ensemble of ISs.

\acknowledgments

Useful discussions are acknowledged with L. Bongini, L. Casetti, L. Delfini, 
S. Lepri, R. Lima, R. Livi, L. Peliti, A. Politi, A. Rampioni, F. Sbrana, L. Tsimiring,
and M. Vassalli. In particular, we are in debt with R. Chelli and P. Procacci
for a critical reading of our results and for fruitful continuous interactions.
Moreover, the field potential here employed has been developed by A. Rampioni 
during his PhD thesis in collaboration with L. Bongini, R. Livi, A. Politi and 
one of the authors (AT).  We acknowledge CINECA in Bologna and INFM for providing
us access to the Beowulf Linux cluster under the
Grant “Iniziativa Calcolo Parallelo”. This work has been partially
supported  by the European Community via the STREP
project EMBIO NEST Contract No. 12835.


\end{document}